\documentclass[sigconf]{aamas}

\usepackage{balance} 
\usepackage{subcaption}
\usepackage{newfloat}
\usepackage{tabularx}
\usepackage{graphicx}

\usepackage{colortbl}
\usepackage{algorithm}
\usepackage{algpseudocode}
\usepackage{amsmath}
\usepackage{enumitem}

\usepackage{amsfonts}

\usepackage{listings}
\DeclareCaptionStyle{ruled}{labelfont=normalfont,labelsep=colon,strut=off}
\floatstyle{ruled}
\newfloat{listing}{tb}{lst}{}
\floatname{listing}{Listing}

\usepackage{amsmath}
\usepackage{booktabs}
\usepackage{float}  

\usepackage{graphicx}  

\usepackage{xurl}

\makeatletter
\gdef\@copyrightpermission{
  \begin{minipage}{0.2\columnwidth}
   \href{https://creativecommons.org/licenses/by/4.0/}{\includegraphics[width=0.90\textwidth]{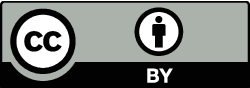}}
  \end{minipage}\hfill
  \begin{minipage}{0.8\columnwidth}
   \href{https://creativecommons.org/licenses/by/4.0/}{This work is licensed under a Creative Commons Attribution International 4.0 License.}
  \end{minipage}
  \vspace{5pt}
}
\makeatother

\setcopyright{ifaamas}
\acmConference[AAMAS '25]{Proc.\@ of the 24th International Conference
on Autonomous Agents and Multiagent Systems (AAMAS 2025)}{May 19 -- 23, 2025}
{Detroit, Michigan, USA}{Y.~Vorobeychik, S.~Das, A.~Nowé  (eds.)}
\copyrightyear{2025}
\acmYear{2025}
\acmDOI{}
\acmPrice{}
\acmISBN{}

\acmSubmissionID{<<845>>}

\title[Ranking Joint Policies in Dynamic Games]{Ranking Joint Policies in Dynamic Games using \newline Evolutionary Dynamics}

\author{Natalia Koliou}
\affiliation{
  \institution{University of Piraeus}
  \city{Piraeus}
  \country{Greece}}
\email{nataliakoliou@iit.demokritos.gr}
\orcid{https://orcid.org/0009-0004-3920-9992}

\author{George A. Vouros}
\affiliation{
  \institution{University of Piraeus}
  \city{Piraeus}
  \country{Greece}}
\email{georgev@unipi.gr}
\orcid{https://orcid.org/0000-0001-5451-622X}

\begin{abstract}

Game-theoretic solution concepts, such as the \emph{Nash equilibrium}, have been key to finding stable joint actions in multi-player games. However, it has been shown that the dynamics of agents' interactions, even in simple two-player games with few strategies, are incapable of reaching \emph{Nash equilibria}, exhibiting complex and unpredictable behavior. Instead, evolutionary approaches can describe the long-term persistence of strategies and filter out transient ones, accounting for the long-term dynamics of agents' interactions. Our goal is to identify agents' joint strategies that result in stable behavior, being resistant to changes, while also accounting for agents' payoffs, in dynamic games. Towards this goal, and building on previous results, this paper proposes transforming dynamic games into their empirical forms by considering agents' strategies instead of agents' actions, and applying the evolutionary methodology \emph{$\alpha$-Rank} to evaluate and rank strategy profiles according to their long-term dynamics. This methodology not only allows us to identify joint strategies that are strong through agents' long-term interactions, but also provides a descriptive, transparent framework regarding the high ranking of these strategies. Experiments report on agents that aim to collaboratively solve a stochastic version of the graph coloring problem. We consider different styles of play as strategies to define the empirical game, and train policies realizing these strategies, using the DQN algorithm. Then we run simulations to generate the payoff matrix required by \emph{$\alpha$-Rank} to rank joint strategies.

\end{abstract}


\newcommand{\BibTeX}{\rm B\kern-.05em{\sc i\kern-.025em b}\kern-.08em\TeX}
\begin{document}
\maketitle
\section{Introduction}
Game theory studies agents' strategies not only in terms of optimality of performance but also with regard to stability of agents' behavior. Game-theoretic solution concepts, particularly the \emph{Nash equilibrium}, have played an important role in this research. However, solution concepts do not account for the long-term dynamics of agents' interactions, which are important in dynamic settings. In static games, where payoff matrices are known, studying solution concepts is relatively straightforward. For example, the mixed strategy \emph{Nash equilibrium} for the Rock-Paper-Scissors game shown in Table~\ref{tab:rps_payoff} occurs when both players randomize their choices uniformly across Rock, Paper, and Scissors. Accounting for the dynamics of agents' interactions over time in dynamic settings, we need to analyze agents' behavior in terms of their payoffs, identifying joint strategies that result into agents' stable behaviors. Evolutionary approaches have shown great potential towards this aim.
\begin{table}[h]
    \centering
    \begin{tabular}{c|c c c}
        & Rock & Paper & Scissors \\ \hline
        Rock     & 0,0    & -1,1   & 1,-1 \\
        Paper    & 1,-1   & 0,0    & -1,1 \\
        Scissors & -1,1   & 1,-1   & 0,0 \\
    \end{tabular}
    \vspace{0.2cm}
    \caption{Payoff matrix for the Rock-Paper-Scissors game.}
    \label{tab:rps_payoff}
\end{table}

To study agents' behavior in multi-agent dynamic settings, researchers often train deep learning models to learn joint policies. These models, either in collaborative or competitive settings, are usually trained with the ultimate objective to result into Nash equilibria, aiming to agents' stability of behavior, where no agent has an incentive to deviate from their joint policy. In complex dynamic settings with long-term dynamics of agents' interactions, there is no guarantee of reaching that objective and there is no way to reveal the reasoning behind the agents' choice of a policy instead of another. Although proposals towards explainability and interpretability of models are important, these aim to provide either explanations for the policy as a whole (i.e. agent's style of play) or about individual decisions. In our case, we need a descriptive framework to account for transparency regarding the strength of agents' policies, accounting for long-term dynamics.

Our goal is to identify agents' individual policies that result in stable (i.e., resistant to changes) behavior while playing with others, accounting for long-term agents' interactions and agents' payoffs, in dynamic games. This is motivated by the need to identify strategies of human or software agents that need to act as co-players in a common setting, considering that these agents have policies that have been formed independently: This is in contrast to assuming that agents have been trained to learn a joint policy. We conjecture that using a descriptive evolutionary framework approach helps agents select strong (i.e. non-transient) policies when playing with or against other agents in dynamic settings. Identifying these policies and understanding their ``superiority'' is important, particularly in collaborative scenarios where agents must choose their own strategies while interacting with others that use specific styles of problem solving.

Towards this goal, we propose exploiting multiple policy models, each realizing a distinct style of play (\emph{strategy}), and then defining an empirical game for evaluating agents' joint performance when they play jointly using various strategy profiles. This empirical game is exploited by the \emph{$\alpha$-Rank} evolutionary framework \cite{omidshafiei2019alpharank}
to evaluate the evolutionary dynamics of agents' strategies over time, ultimately identifying which ones prevail in the long run. 

Although this work builds on the \emph{$\alpha$-Rank} framework, it contributes a perspective for evaluating individual agents' strategies in stochastic, sequential decision making settings, when they act with other agents following specific styles of play.\footnote{The source code is publicly available at: \url{https://github.com/nataliakoliou/Collaborative-Graph-Coloring-Game}.} In so doing, we:
\begin{itemize}
    \item Describe a concise methodology for evaluating and ranking agents' policies, accounting for their long-term interactions in dynamic settings, using the \emph{$\alpha$-Rank} evolutionary framework.
    \item Demonstrate this methodology in multi-agent graph coloring dynamic games, defining multiple styles of play (i.e. strategies) per agent.
    \item Show how agents' choices of strategies—and the policies realizing them—can be transparently justified by means of a descriptive framework.
\end{itemize}


\section{Background}

In this section, we outline the key concepts necessary to follow the proposed approach. 

\subsection{Dynamic Games}

Dynamic games describe agent interactions along the time dimension. Unlike static games, where players execute single one-shot actions, dynamic games involve a series of decisions made by each of the players at subsequent points in time. A key property of dynamic games is that the actions taken at any given moment influence the future states of the system and future decisions made. These temporal dependencies require players to consider the long-term consequences of their actions. A dynamic game can be represented as a tuple 
$G = (S, K, A, T, P) $, 
where $S$ represents a finite set of states, $K$ is the set of players, and $A = (A^k \times A^{-k})$ is the set of joint actions, where $A^k$ corresponds to the set of actions available to the player $k$. $A^{-k}$ denotes the set of actions available to players other than $k$. The transition function $T$ describes the dynamics of the setting, determining the next state of the system based on the current state and the actions chosen by the players. Finally, $P^k: S \times (A^k \times A^{-k}) \times S \to \mathbb{R}$ is the payoff function for player $k$, given the current joint state, the action chosen by player $k$ and the actions of the other agents, and the resulting state. 

In this work we focus on stochastic dynamic games, as introduced by L.S. Shapley in 1953 \cite{Shapley1953StochasticG}. In stochastic games, the outcome of players' actions is influenced by probabilistic events, making future states of the game uncertain. These games are often referred to as Markov games \cite{Shoham_Leyton-Brown_2008}. Therefore, in stochastic games, the transition function $T$ is defined as a probability distribution over next states. Specifically, $T: S \times A \to \Delta(S)$, where $\Delta(S)$ is a probability distribution over the states, given a state and joint action. For example, in poker, while players' actions do influence the outcome, the next state of the game also depends on luck, such as drawing a strong hand like a flush or a weak hand like a pair of twos. In such games, players, when planning their actions, must account for both the actions of their opponents and the dynamics of the environment.

In dynamic games, players aim to decide on the course of their joint actions through time (joint policy) to maximize their accumulated rewards over time: 

\[\sum_t T(s_t, (a_t^k, a_t^{-k}), s_{t+1}) \cdot P^k(s_{t}, (a_t^k, a_t^{-k}),s_{t+1})\]
%
Here, $T$ specifies the transition probability from state $s_t$ to the state $s_{t+1}$ given $(a_t^k, a_t^{-k})$, and $P^k(s_{t}, (a^k, a^{-k}), s_{t+1})$ is the reward the player receives for choosing action $a^k$, given the actions $a^{-k}$ of the other players, at state $s_{t}$, and resulting into state $s_{t+1}$. 

\subsection{Empirical Analysis and Empirical Games}

Empirical Game Theory Analysis (EGTA) provides a framework that uses empirical methods to analyze player interactions within complex game environments \cite{Levet2016GameT}. These methods are used to define game components, such as payoff matrices, based on observed interactions, rather than relying on predefined rules. Simulation is one such method, where agents repeatedly play a game, and payoffs are collected based on the outcomes of these interactions. Other techniques include sampling, where a subset of the action space is explored to approximate the payoffs for a wider set of actions, and machine learning methods to identify players' behavior and estimate outcomes based on historical data \cite{wellman2024empiricalgametheoreticanalysissurvey}. Empirical techniques are applied in cases where the action space is too large and complex to define manually, making payoff matrices impossible to generate from simple rules and assumptions.

An empirical game, also referred as a meta-game, is a \emph{Normal Form Game} of the form 
$G = (K, \mathcal{S}tr, P) \label{eq:nfg}$, 
where $K$, $\mathcal{S}tr$ and $P$ specify players, players' strategies, and payoffs, correspondingly. We define empirical games by abstracting the actions and defining the payoffs of players in an underlying dynamic game. The underlying game represents the actual setting where players interact. In the empirical game representation, $K$ is the same as in the underlying game, i.e. the set of players engaged in strategic interactions. 
Strategies (i.e. styles of playing the game) in empirical games offer an action abstraction and can be derived by identifying distinct behaviors during game-play. The strategy space $\mathcal{S}tr$ consists of distinct agents' styles of play. $\mathcal{S}tr^k$ denotes the strategies of agent $k$ and $\mathcal{S}tr^{-k}$ the set of strategies of agents other than $k$. The set of strategy profiles, i.e. agents' joint strategies, is defined to be $\mathcal{SP}=\{\mathcal{S}_i | \mathcal{S}_i = (str_i^1, str_i^2, \allowbreak \dots, str_i^K)\} \text{, where } str_i^k \in \mathcal{S}tr^k\text{, and }i=1,\dots\ \text{the profile index}\}$.
The payoff matrix of an empirical game can be generated using empirical analysis techniques. Here, we focus on simulation, where agents engaged in the underlying game act according to policies adhering to specific strategies. 

Subsequently, we use the terms \emph{action} and \emph{policy} when speaking about the underlying game, and the term \emph{strategies} or \emph{styles of play} when speaking about the empirical game. 

The payoff function $P$ of the empirical game is computed from simulations for each strategy profile as follows:

\begin{eqnarray}
P^k(str^k, str^{-k}) =\frac{1}{N} \cdot \sum_{i=1}^{N} P^k_{i}(str^k, str^{-k}) \nonumber
\label{eq:meta_payoff}
\end{eqnarray}

where $N$ is the number of simulation runs, $str^k$ represents player $k$'s strategy, $str^{-k}$ denotes the strategies of the other players, and $P^k_{i}(str^k, str^{-k})$ (with an abuse of notation) represents the payoff player $k$ receives in simulation run $i$ when playing strategy $str^k$ against the strategies of the other players. It must be noted that in contrast to dynamic games the payoff function does not take states as arguments, as the outcomes are determined by agents' joint strategies, i.e. $P^k: (\mathcal{S}tr^k \times \mathcal{S}tr^{-k}) \to \mathbb{R}$ \cite{omidshafiei2019alpharank}. If we aggregate these expected payoffs into a matrix, we get the empirical payoff matrix whose dimensionality is $\prod_{k=1}^K\mathcal{S}tr^k$. Each entry represents the expected payoff for strategy $str^k$ against strategy $str^{-k}$.

\subsection{The \texorpdfstring{$\alpha$}{alpha}-Rank Method}

Evolutionary dynamics studies how agents' interactions in multi-agent settings evolve over time. While single-agent systems have acquired a strong foundation over the years \cite{10.5555/2831071.2831085}, multi-agent systems are more challenging to analyze.

Current literature indicates a growing interest in studying the evolutionary dynamics of multi-agent systems \cite{10.5555/2831071.2831085} \cite{David_2014} \cite{paul2022multiagentpathfinding}. 
In the context of games, evolutionary algorithms are widely used to explore game-theoretic concepts, resulting to the \emph{Evolutionary Game Theory}. Building on work done in this area, \emph{$\alpha$-Rank} \cite{omidshafiei2019alpharank} introduces a novel game-theoretic approach to provide insights into the long-term dynamics of agents' interactions. 

\emph{$\alpha$-Rank} is an evolutionary methodology designed to evaluate and rank agents' strategies in large-scale multi-agent interactions, introducing a new dynamic solution concept called \emph{Markov-Conley chains} (MCCs). Given a K-player game, \emph{$\alpha$-Rank} considers the empirical game with K player slots, called $populations$, where individual agents correspond to strategies, i.e. to styles of playing the underlying game. 
Populations of agents interact with each other through an evolutionary process following the dynamics of games. The rewards received from these interactions determine how well each strategy performs and, in turn, how often it is adopted by individuals in the populations. Strategies that perform well have a higher probability of being adopted and carried over to the next generation, while those performing poorly are less likely to be adopted. This process leads to the evolution of populations. 

To facilitate evolution, \emph{$\alpha$-Rank} uses the concept of mutation. Initially, populations are monomorphic, meaning all individuals within them choose the same strategy. During K-wise interactions, individuals have a small probability of mutating into different strategies or choosing to stick with their current one. The probability that the mutant will take over the population, defined to be the fixation probability function $\rho$, depends on the relative fitness of the mutant and the population being invaded. Fitness is a function that computes the expected reward an individual can receive when adopting a particular strategy, given the strategies of the other individuals. The stronger the fitness, the more likely it is for individuals to mutate, whereas the lower the fitness, the more likely it is for the mutant to go extinct. When the mutation rate is small, we can assume that the fitness for any agent $k$ is $f^k(str^{k}, str^{-k}) = P^k(str^{k}, str^{-k})$, where $P$ is the empirical game payoff.

Formally, the probability of a mutant strategy $str'$ fixating in some population where individuals play strategy $str$ is given by:
\begin{eqnarray}
\rho_{str \to str'} = \frac{1 - e^{-\alpha \cdot \Delta f}}{1 - e^{-\alpha \cdot m \cdot \Delta f}} \label{eq:fixation_prob}
\end{eqnarray}
assuming that $\Delta f$ is non-zero. $\Delta f = f^k(str', str^{-k}) - f^k(str, str^{-k})$ represents the difference in fitness between the mutant strategy $str'$ and the resident strategy $str$ in the focal population $k$, while the remaining $K - 1$ populations are fixed in their monomorphic strategies $str^{-k}$. Parameter $m$ is the population size and $\alpha$ is the selection intensity. This adjusts the sensitivity of the system to fitness differences: with higher values of $\alpha$, even small differences in fitness lead to larger changes in $\rho$. The nominator measures the potential of the mutant to ``invade'' the resident population solely based on its fitness advantage. Note that, for example, as $\Delta f$ approaches zero, the probability of the mutant's success decreases. The denominator, on the other hand, normalizes the fixation probability using the population size $m$, making it more challenging for a mutant to dominate in larger populations. When $\Delta f$ is zero, the fixation probability is equal to $1/m$ (\cite{omidshafiei2019alpharank}, eq.13), indicating that the mutant strategy has the same probability of taking over as any other strategy in the population. We refer to this probability as the \emph{neutral fixation probability}, denoted by $\rho_m$. \\
In the context of K-player games, \emph{$\alpha$-Rank} creates a Markov transition matrix over strategy profiles. This is an $|\mathcal{S}tr| \times |\mathcal{S}tr|$ matrix that defines the probability of moving from one strategy profile to another based on how likely each population is to change its strategy.
\begin{eqnarray}
C_{str \to str'} = 
\begin{cases} 
\eta \cdot \rho_{str \to str'} & \text{if } str \neq str' \\ 
1 - \sum_{str \neq str'} C_{str \to str'} & \text{otherwise}
\end{cases} \label{eq:transition_matrix_entry}
\end{eqnarray}
Here, $C$ is the strategy-transition matrix where each entry $C_{str \to str'}$ represents the probability of transitioning from strategy $str$ to strategy $str'$. The first part of the formula calculates the probability of strategy transition, $\rho_{str \to str'}$ scaled by $\eta = \frac{1}{\sum_{k} (|Str^k| - 1)}$, where $k$ indexes populations. The second part of the formula computes the probability of staying with the same strategy $str$, excluding transitions to all other strategies. \\
This evolutionary process of competition and selection among players' strategies leads to a unique stationary probability distribution $\pi$ of dimensionality $|\mathcal{SP}|$, where the mass assigned to a strategy profile indicates how likely it is to resist being ``invaded'' by other strategies as the dynamics evolve. To evaluate and rank strategy profiles —which is the ultimate goal— the method calculates $\pi$ over the game's Markov chain, using the strategy-transition matrix $C$. This distribution indicates how often the system is likely to remain in each profile over time, allowing us to identify the most dominant strategies that are expected to prevail in the long run. Formally, $\pi$ can be computed from the following equation:

\begin{equation}
\pi C = \pi \Rightarrow \pi (C - \mathbb{I}) = 0 \label{eq:stationary_distribution}
\end{equation}

where $\mathbb{I}$ is the identity matrix. This means we are looking for a probability vector $\pi$ such that when multiplied by the transition matrix $C$, it remains unchanged. To solve for $\pi$, the augmented matrix from $C - \mathbb{I}$ is constructed and a normalization condition to ensure that probabilities sum to 1 is imposed\footnote{The system $\pi(C - \mathbb{I}) = 0$ by itself does not have a unique solution, as there are infinitely many vectors $\pi$ that satisfy it. To get a unique solution $\pi=(\pi_1,\pi_2,..., \pi_{|\mathcal(SP)|})$, it must hold that $\sum_{i} \pi_i = 1$.}. In this stationary distribution, $\pi=(\pi_1,\pi_2,..., \pi_{|\mathcal(SP)|})$, each $\pi_i$ represents the average time the system spends in strategy profile $i$.


\section{Problem Statement}

As already stated, we aim at identifying (human and software) agents’ strong joint
strategies, in terms of stability and joint performance, to solve problems in dynamic settings, accounting for agents' long-term dynamics of interactions. Stability implies non-transient strong strategies, persisting in time, as they fit better to the objective of the agents given the structure of the game and payoffs received. However, in dynamic games, we need to define the payoff matrix and exploit this to determine strategies stability. Even if we manage to estimate payoffs, the computation of solution concepts like the \emph{Nash equilibrium} imposes a high computational cost in these settings, does not guarantee convergence, and fails to scale to large games. Beyond identifying stable joint strategies, it is important to transparently justify/describe what makes one joint strategy better than another, providing evidence for the rankings. 

We could, therefore, consider our problem as follows: Given a dynamic game $G$ with $K$ players, our goal is to identify styles of playing $G$, and thus, the set of strategy profiles $\mathcal{SP}$, and rank these profiles based on how stable they are over time, considering long-term agents' interactions towards achieving their objectives. Specifically, we aim to define a ranking function $\mathcal{R}: \mathcal{SP} \to \mathbb{R}$, where $\mathcal{R}(\mathcal{S}_i) > \mathcal{R}(\mathcal{S}_j)$ (resp. $\mathcal{R}(\mathcal{S}_i) \geq \mathcal{R}(\mathcal{S}_j)$) indicates that the strategy profile $\mathcal{S}_i$ is strictly (resp. weakly) preferred over $\mathcal{S}_j$, using a descriptive framework $\mathcal{D}$ defined over $\mathcal{SP}$, that provides transparency on how rankings are decided. 

It must be noted that empirical game strategies are realized by agents' policies adhering to these strategies in the underlying game. Thus, identifying stable joint strategies in the empirical game translates to identifying stable joint policies adhering to these strategies in the underlying dynamic game.


\section{Proposed Method}

To address the challenge of identifying stable joint policies in dynamic games, we propose an approach that combines concepts from \emph{Empirical Game Theory} and \emph{Evolutionary Dynamics}, using \emph{$\alpha$-Rank}, providing transparency to rankings of agent's styles of play.

Given that the set of agents' policies in dynamic games can be infinitely large we focus on a subset of policies that adhere to concrete and well-defined styles of play. A way to identify styles of play is to observe how players behave in the underlying game or exploit demonstrations of game playing by means of style (mode) - preserving offline or inverse reinforcement learning methods. This may result into a mixture of policies (one per style of play) given that human experts performing a task usually follow a distinct set of specific styles based on well-established practices, preferences and experience. Having determined the game playing strategies, we can transform the dynamic game into its empirical form, defining the meta-game, as specified in Section 2.2: By (a) identifying empirical game strategies, and (b) training policies for agents to play the underlying game according to these strategies, (c) run policies to define the empirical game payoff matrix, through simulations.

Having defined the meta-game, we need to define the function $\mathcal{R}$, which ranks joint strategies based on agents' long-term dynamics and objectives. In our approach, we propose using the evolutionary \emph{$\alpha$-Rank} methodology to determine these rankings. The rankings are based on each strategy profile's evolutionary success, which is reflected in the probability of that profile being selected over time. This probability is captured by the stationary distribution $\pi$, which \emph{$\alpha$-Rank} computes in the limit of infinite ranking intensity $\alpha$. As demonstrated by \cite{omidshafiei2019alpharank}, a large $\alpha$ limit suffices. Therefore the long-term behavior is captured by the unique stationary distribution $\pi$ under the large $\alpha$ limit. 
As it is proved in \cite{omidshafiei2019alpharank}, the Markov chain associated with a generalized multi-population model, coincides with the MCC solution concept: MCCs can be identified effficiently in all games by the sink strongly connected components of a response graph, whose vertices correspond to pure strategies' profiles and directed edges from a strategy profile $\mathcal{S}_i$ to a strategy profile $\mathcal{S}_j$ specifies that $\mathcal{S}_j$ is weakly a better response than $\mathcal{S}_i$ for player $k$.

To compute $\pi$ over strategy profiles, \emph{$\alpha$-Rank} requires the payoff matrix of the empirical game $P$. Along with the stationary distribution $\pi$, \emph{$\alpha$-Rank} outputs the fixation probability function $\rho_{\mathcal{S}_i \to \mathcal{S}_j}$, $\mathcal{S}_i, \mathcal{S}_j \in \mathcal{SP}$, which measures the likelihood of transitioning from one strategy profile $\mathcal{S}_i$ to another $\mathcal{S}_j$. Thus, \emph{$\alpha$-Rank} can be abstracted as a function:

\begin{equation}
\alpha\text{-Rank}(P) \rightarrow (\pi, \rho) \label{eq:abstract_arank}
\end{equation}

While the stationary distribution $\pi$ provides valuable insight into the long-term behavior of strategies, it alone does not help us fully understand how strategies transition between one another. The fixation probability function $\rho$ fills this gap. Based on this, the descriptive framework $\mathcal{D}$ can be adequately represented by $\pi$ and $\rho$, which are constituents of the response graph that provides a complete view of the empirical game dynamics.

Overall, building on the \emph{$\alpha$-Rank} descriptive framework, the method proposed here for computing strategy profile rankings in dynamic games is as follows:

\begin{enumerate}[itemsep=0.25em]
  {\fontsize{8.5}{10}\selectfont
      \item Identify  players' styles of play.
      \item Define the strategies of the empirical game based on those styles.
      \item Train policies realizing the defined strategies. 
      \item Run game simulations to create the empirical payoff matrix $P$.
      \item Apply \emph{$\alpha$-Rank} to define $\mathcal{R}$ and $\mathcal{D}$:
      \begin{enumerate}[itemsep=0.2em]
          \item Calculate the Markov transition matrix $C$.
          \item Find the unique stationary distribution $\pi$.
          \item Rank joint strategies by ordering the masses of $\pi$.
          \item Describe the rankings through the response graph.
          \item Study the effect of different $\alpha$ values on $\pi$.
      \end{enumerate}
  }
\end{enumerate}


\section{Experiments and Results}

In this section, we present the experiments and discuss the results obtained from applying the proposed methodology to the \emph{Graph Coloring Game} using the configuration shown in Figure~\ref{fig:grid_and_graph}. Appendix $C$ provides results with additional grid configurations.

\subsection{The Graph Coloring Problem}

The well-known \emph{Graph Coloring Problem} (GCP) involves assigning colors to vertices in a graph such that no two adjacent vertices share the same color, and using the minimum number of colors, also known as the chromatic number \cite{watkins2023generating}.

In this study, we shift our focus from finding the chromatic number across graph configurations to solving the multi-agent problem of assigning colors to vertices of a dynamic graph with respect to the constraints: In doing so, we define the graph coloring problem as a dynamic game that allows us to study the evolutionary dynamics in multi-agent interactions. 
This problem setting abstracts settings where agents need to abide to dynamically evolving constraints while acting jointly. For instance, in traffic, any agent acts to abide to constraints, and these actions, given also the dynamics of the environment, result into new emerging constraints to which agents must adhere to, and so on. Through this problem, we aim to demonstrate how we can gain insights into the effectiveness of playing the dynamic game when individual styles of play are combined. 


%
\begin{figure}[h]
 \centering
  \begin{minipage}{0.42\linewidth}
    \centering
    \includegraphics[width=\linewidth]{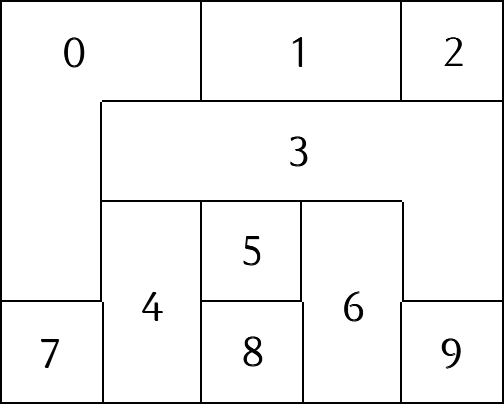}
    \Description{The $4 \times 5$ grid environment used for our graph coloring game.}
  \end{minipage}
  \hspace{0.03\textwidth}
  \begin{minipage}{0.42\linewidth}
    \centering
    \includegraphics[width=\linewidth]{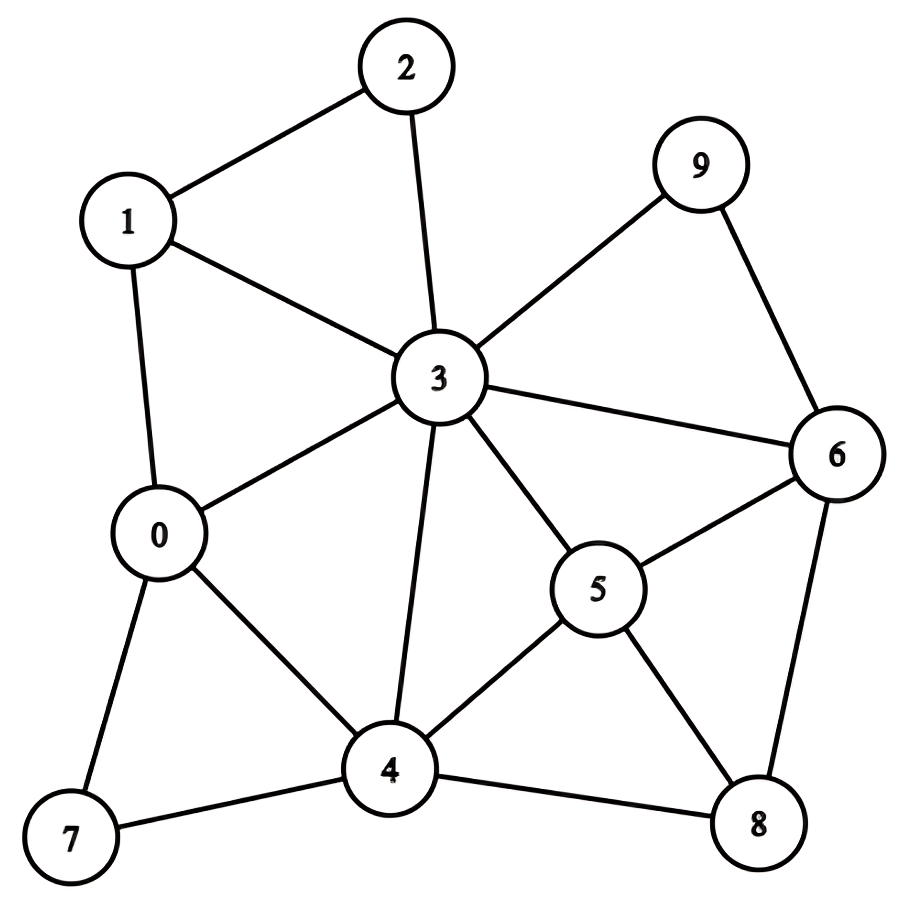}
    \Description{Graph representation of the grid environment.}
  \end{minipage}
  \caption{A snapshot of the game environment in grid and graph forms.}
  \label{fig:grid_and_graph}
  \Description{A snapshot of the game environment in grid and graph forms.}
\end{figure}

We consider the underlying graph coloring dynamic game to be a two-player game executed in rounds. The graph corresponds to a grid comprising blocks of cells: A block comprises one or more merged cells. Each vertex of the graph corresponds to a block, and the adjacency relation between blocks specifies the edges in the graph. At the beginning of the game, the grid is initialized with a random number of rows and columns ($n \times m$). In our experimental setup we assume a $4 \times 5$ grid. The environment is initialized by randomly combining cells to create the blocks. The resulting configuration remains the same throughout the entire game. A snapshot of such a configuration with 10 blocks is the one shown in Figure~\ref{fig:grid_and_graph}, together with the corresponding graph. Merging cells is important as it allows for complex neighboring relationships to be defined, expanding beyond the standard constraints between adjacent blocks. Blocks are either (a) colored by the agents, (b) white (free to be colored), or (c) hidden (their colors cannot be observed and they cannot be recolored by the agents). Let $B$ be the set of blocks corresponding to graph vertices and $CR$ be the set of possible colors that an agent can use for coloring blocks in $B$. The game unfolds over multiple rounds in which agents choose their actions simultaneously. At the beginning of each round, the environment reveals the color of some of the hidden blocks, if any. The number of blocks that get unhidden is random, which implies that the state of the graph is influenced not only by the agents' actions but also by the environment. We therefore consider the game to be stochastic. When all blocks in $B$ are uncovered and colored, the game ends. 

The set of agents' actions $A$ is defined to be the Cartesian product of the set of blocks $B$ and the set of the available colors $CR$:

\begin{eqnarray}
A = B \times CR = \{(b, c) \mid b \in B, \, c \in CR\} \label{eq:action_space}
\end{eqnarray}

To specify states, let $CR^*$ include the elements of $CR$, and two additional elements representing hidden and white blocks: $CR^* = CR \cup \{\text{hidden}, \text{white}\}$. A state $s$ is as follows:

\begin{align*}
s = \{(b_i, c_i), i=1, \dots, |B|\}, \\ s.t. \forall b \in B, \, \exists \text{ a unique } c \in CR^*\text{, with } (b,c) \in s
\end{align*}

Regarding the reward function, it is a sum of gains, penalties, sanctions, delays and adopted preferences. Given that actions are represented as vectors of shape $(b, c) \in B \times CR$, an agent receives a gain point (+1) for each neighbor of $b$ that has a different color than the chosen color $c$. On the contrary, an agent receives a penalty point (-2) for each neighbor that shares the same color $c$. Sanction is a big negative reward (-10) that an agent receives when it attempts to color a hidden block or a block that has already been colored. Delay (-1) is a small negative reward that both agents receive when they try to color the same block $b$, causing a brief pause in the game to determine which agent will eventually color $b$. Last but not least, there is the preference-adoption reward, which agents receive regardless of whether their action is good, bad, or forbidden. This reward helps agents to be trained so as to adhere to specific preferences, or what we call \emph{styles of play}. We will elaborate shortly on these in the following section.

\subsection{Defining the Empirical Game}

Transforming the underlying dynamic graph-coloring game into its empirical form involves two key steps: (1) identifying agents' strategies and (2) constructing the empirical game payoff matrix.

\subsubsection{Agents' Strategies}
\label{sec:5.2.1}

Agents' strategies define distinct styles of play, usually revealed by preferences in playing the game. In our experiments, we specify different styles across three main dimensions: color tone (preference for which colors to use), block difficulty (preference for the types of blocks to choose), and coloring approach (preference for the number of colors to use), as shown in Table~\ref{tab:preferences}. Through the combination of preferences in each of these dimensions, a style can range from complete indifference, where none of the dimensions hold any influence (denoted by ``I''), to specific preferences in all dimensions.

Policies corresponding to specific strategies are represented using convolutional neural networks. To train these policies, we assign specific values in the three dimensions of the game's \emph{preference} reward. These values, range from -1 to 1, where 1 indicates a strong preference for a particular dimension. For example, a value of 0.7 for warm colors indicates a relatively high preference for warm tones. In our experimental setting, we define 11 distinct styles: I, C, W, E, M, L, A, AE, CA, LE and WL, given ``I'' and combinations of preference values specified in Table \ref{tab:preferences}. Assuming no inherent bias among the empirical game players, we allow populations to sample from the same list of strategies.
\begin{table}[h]
    \centering
    \begin{tabular}{cc}
        \toprule
        \textit{Preference Dimension} & \textit{Value}\\ 
        \midrule
        Color Tone & warm (W) vs. cool (C) \\ 
        Block Coloring Difficulty & small (L) vs. large (A)\\ 
        Coloring Approach & minimalistic (M) vs. extravagant (E)\\ 
        \bottomrule
    \end{tabular}
    \caption{Dimensions specifying agents' strategies}
    \label{tab:preferences}
    \Description{bla bla}
\end{table}

\subsubsection{Training the agents}

\begin{figure}[h]
  \centering
  \includegraphics[width=1\linewidth]{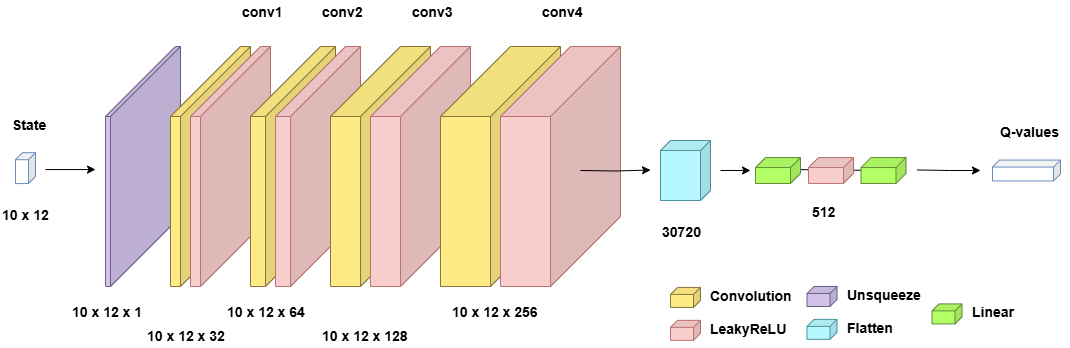}
  \caption{Convolution Policy Network Architecture}
  \label{fig:convDQN}
  \Description{A diagram of the ConvDQN architecture, showing the sequence of convolutional layers, activation functions, and fully connected layers.}
\end{figure}

All policy models share the same underlying architecture and training setup. Although hyperparameter tuning is typically recommended, it does not make much difference in this case, as these models are relatively easy to optimize when trained in small settings. Regarding the convolutional neural network architecture, it consists of four convolutional layers, each defined with a kernel size of 3, stride of 1, and padding of 1, meant to extract spatial features from the input. The input tensor has dimensions $10 \times 12$, where $|B|=10$ represents the number of blocks in the state and $|CR^*|=12$ represents the number of possible colors a block can have. Each block is encoded using one-hot encoding, meaning that each color is represented as a binary vector of length 12. As shown in Figure \ref{fig:convDQN}, the network processes the input through four convolutional layers with respective output sizes 10 × 12 × 32, 10 × 12 × 64, 10 × 12 × 128, and 10 × 12 × 256. The final convolutional layer output is flattened to a vector of size 30720. This representation is passed through two fully connected layers: the first reduces the size to 512, and the second produces the final output, which corresponds to a matrix of size $|CR| \times |B|= 10 \times 10$ encoding the action space. Although the architecture seems to be excessive, it aims at efficient policy training and generalization, by extracting meaningful relationships among state components. This claim is justified in Appendix D.

Policy models are trained individually (without co-players) in the underlying game using the deep Q-learning reinforcement learning algorithm specified in Algorithm 1. We set $\gamma$ to 0.7. To optimize the model parameters, we use the smooth L1 loss function with $\beta$=1.0 and the Adam optimizer with a learning rate of 5e-4 and weight decay of 1e-5 to prevent over-fitting. To further enhance the learning process, we incorporate experience replay, with a memory that stores up to 10 million experiences \cite{10.1007/BF00992699}. A target network alongside the main policy network, is being used according to the Double-DQN approach \cite{vanhasselt2015deep}. To update the target network we apply a soft update with a factor $\tau$=5e-3. This gradually brings the target network closer to the policy network, balancing learning speed and stability. With a batch size of 64, we train the models for 10000 episodes. All trained agents manage to learn a policy that successfully abides to constraints.

\begin{algorithm}
\caption{Double Deep Q-Learning with Experience Replay}
\begin{algorithmic}[1]
\State $Q_\theta$, $Q_{\theta'} \gets Q_\theta$, $M$ \Comment{Initialize policy/target nets \& memory}
\For{episode}
    \State $s \gets s_{0}$
    \For{step}
        \State $a \gets \text{argmax}_a Q_\theta(s)$ \Comment{Select $\epsilon$-greedy action}
        \State $(s, a, r, s') \in M$ \Comment{Store experience}
        \If{$|M| > \text{batch size}$} 
            \For{each $(s, a, r, s')$ in $M$} \Comment{Sample memory}
                \State $y \gets r + \gamma \max_{a'} Q_{\theta'}(s')$
                \State $L \gets \text{Loss}(Q_\theta(s), y)$
                \State $\theta \gets \theta - \alpha \nabla_\theta L$
            \EndFor
        \EndIf
        \State $Q_{\theta'} \gets \tau Q_\theta + (1 - \tau) Q_{\theta'}$ \Comment{Soft update}
        \State $s \gets s'$
    \EndFor
\EndFor
\end{algorithmic}
\end{algorithm}

\subsubsection{Empirical Game Payoff Matrix}

The empirical payoff matrix is generated by simulating each strategy profile over multiple games. Payoffs represent how well different styles of play perform jointly, according to the game’s rules. Here we must note that even the most incompatible pairs of styles violate few constraints, due to the inherent differences between agents’ preferences. For instance, the profile (W,C) of compatible strategies scores 0.05\% of violations, while the profile of incompatible strategies (C,C) scores 23\% violations. 

The values in the payoff matrix are computed in terms of the delay and the quality of the solution according to the game's constraints (gain, penalty, and sanction), excluding preferences. This ensures a common ground for distinct strategies, evaluating solutions solely based on the game's rules. For each pair of strategies, we simulate the game over 5,000 repeats and calculate the average payoff for each strategy. These values are then organized into the payoff matrix, which is provided in Table~\ref{tab:gcg_payoff_matrix} (\hyperref[appendix:A]{Appendix A}). From this matrix, we observe that (L, WL) and its symmetric counterpart (WL, L) both with payoffs of (3.15, 3.21) and (3.21, 3.15) respectively, are the only Nash equilibria. It is important to note here that these equilibria prescribe agents' strategies, given that they do play the game with rational co-players, but they do not capture the overall dynamics of the game, considering the long-term effects of agents' interactions.

\subsection{Evaluation and Ranking}

Given the payoff matrix derived from the empirical analysis, we apply the \emph{$\alpha$-Rank} method to evaluate the performance of strategy profiles over time in terms of the MCC solution concept. Specifically, we ran the method 1000 times, using values of $\alpha$ within the range $[0.1, 10]$ with step=0.01, while assuming populations of size $m=100$. We provide as input the strategies defined in Section~\ref{sec:5.2.1} and the empirical game payoff matrix. We focus on the rankings of the top 6 strategy profiles, to identify the stronger ones across different values of $\alpha$.

As we observe from the rankings in Table~\ref{tab:ranking_table}, the strategy profile that prevails in the long run is (WL, CA); this is the primary component of the MCC. Although the table was derived using an $\alpha$ value of 2, the rankings remain consistent even when $\alpha$ is set to 10. We choose $\alpha=2$ over $\alpha=10$, to display the rankings of lower-performing strategy profiles, which would otherwise drop to zero. First, it is worth mentioning that the Nash equilibria (L, WL) and (WL, L) don't appear among the top-ranked strategy profiles. This is because MCC components are defined based on how well strategies perform when interacting with other strategies, based on long-term agents interactions. The individual strategies within the Nash equilibrium profile, either WL or L, may not result in favorable interactions with other strategies. As a result, the profile (WL, L) is ranked lower than others.

To further support our observations regarding the misalignment between the two solution concepts, let's examine why (CA, WL) is part of the MCCs, while (L, WL), the Nash equilibrium, is not. A closer look at the payoff matrix in Table~\ref{tab:gcg_payoff_matrix} reveals that L appears to be the worst-performing strategy for the row player, with an average payoff of 3.13. In this case, being in the Nash equilibrium means the player is stuck with a strategy that gives low rewards, making it the best among other options, rather than a strong choice. If it happens to play this strategy, it would expect its rational opponent to play WL. Strategy CA on the other hand, is the best-performing strategy for the row player, with an average payoff of 3.18. Combined with WL, which is the best performing strategy for the column player, with an average payoff of 3.18, make (CA, WL) to be the top ranked strategy profile in the ranking Table~\ref{tab:ranking_table}. 

\begin{table}[h]
    \centering
    \begin{tabular}{lcc}
        \hline
        \textbf{Profile} & \textbf{Rank} & \textbf{Score} \\
        \hline
        (WL, CA) & 1 & 0.42 \\
        (W, CA) & 2 & 0.13 \\
        (M, CA) & 3 & 0.12 \\
        (CA, M) & 4 & 0.08 \\
        (CA, W) & 5 & 0.08 \\
        (CA, LE) & 6 & 0.01 \\
        \hline
    \end{tabular}
    \caption{Strategy profiles' rankings for $\alpha=2$}
    \label{tab:ranking_table}
\end{table}

Rankings within the MCC are also very intuitive. For example, strategies that prefer different color tones, such as (WL, CA) or (W, CA), tend to result into fewer conflicts since, they naturally avoid selecting the same colors. Similarly, strategies that prefer different blocks based on their difficulty, such as (WL, CA) or (CA, LE), tend to provide solutions with minimal delay, as they naturally avoid coloring the same blocks. Notably, profiles with mixed preferences across these dimensions demonstrate the most promising performance, which explains why (WL, CA), as such a profile, is a key component of the MCC. However, not all profile rankings can be easily explained through the game's rules alone; the expected influence of certain strategies on the quality of the solutions remains ambiguous. For example, profiles with strategies like M and E are more difficult to analyze.

\begin{figure}[h]
  \centering
  \includegraphics[width=0.9\linewidth]{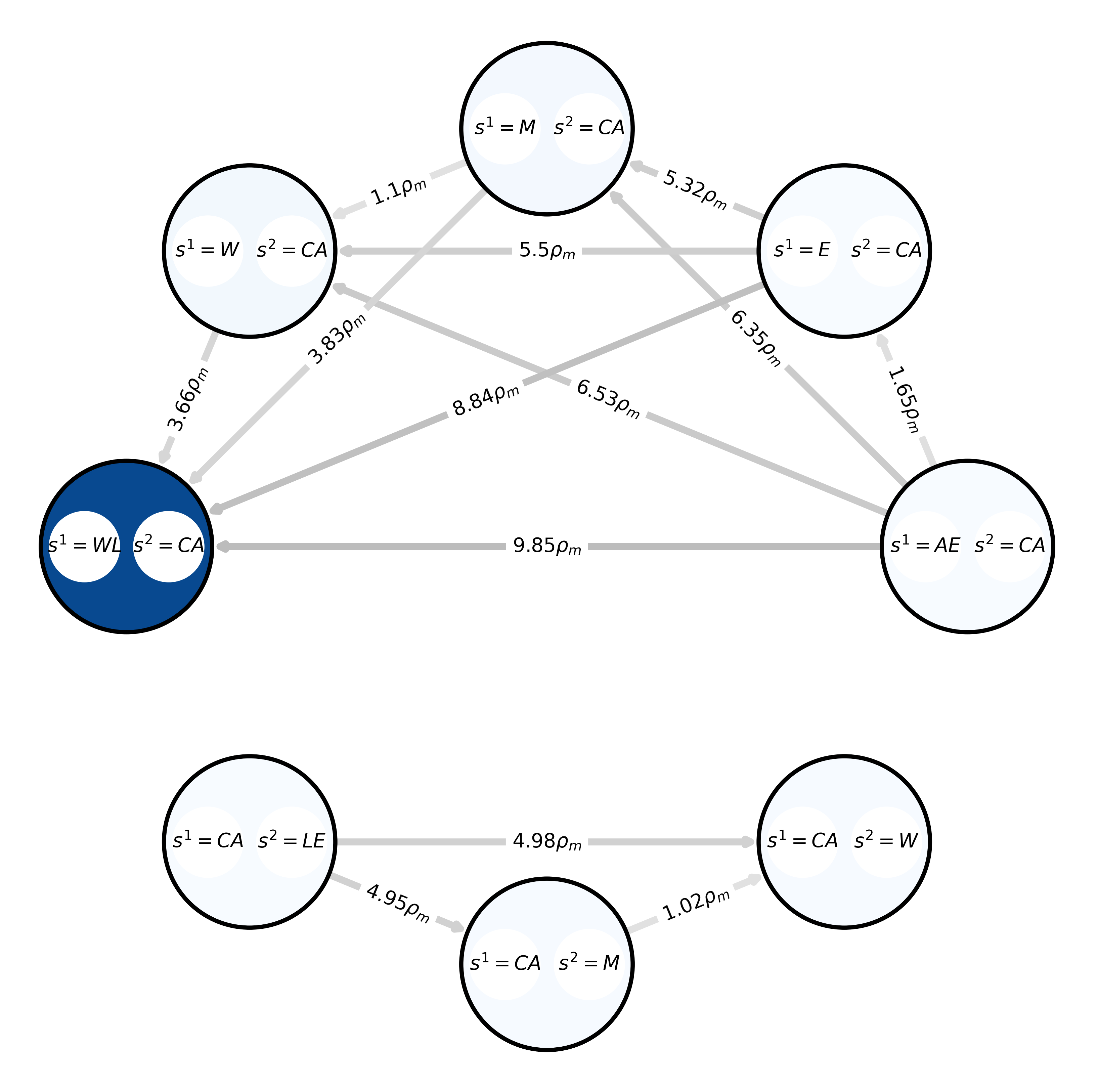}
  \caption{Response graph for $\alpha=6.4$.}
  \label{fig:response_graph_6.4}
  \Description{Response graph for $\alpha=6.4$.}
\end{figure}

The response graph provides a visualization to interpret the \emph{$\alpha$-Rank} results. This graph illustrates the MCC, using the strategy profiles' masses from the stationary distribution, $\pi$, along with the fixation probability function $\rho$ provided by \emph{$\alpha$-Rank}. Figure~\ref{fig:response_graph_6.4} shows the response graph for $\alpha = 6.4$. We consider it to be part of the descriptive framework $\mathcal{D}$, as it offers insights into how rankings were derived. Additional graphs for $\alpha = 0.4$, $1.3$, and $1.9$ are available in Figure~\ref{fig:response_graphs_appendix} (\hyperref[appendix:B]{Appendix B}).

\begin{figure*}[h]
  \centering
  \begin{subfigure}[b]{0.49\linewidth}
    \includegraphics[width=\linewidth]{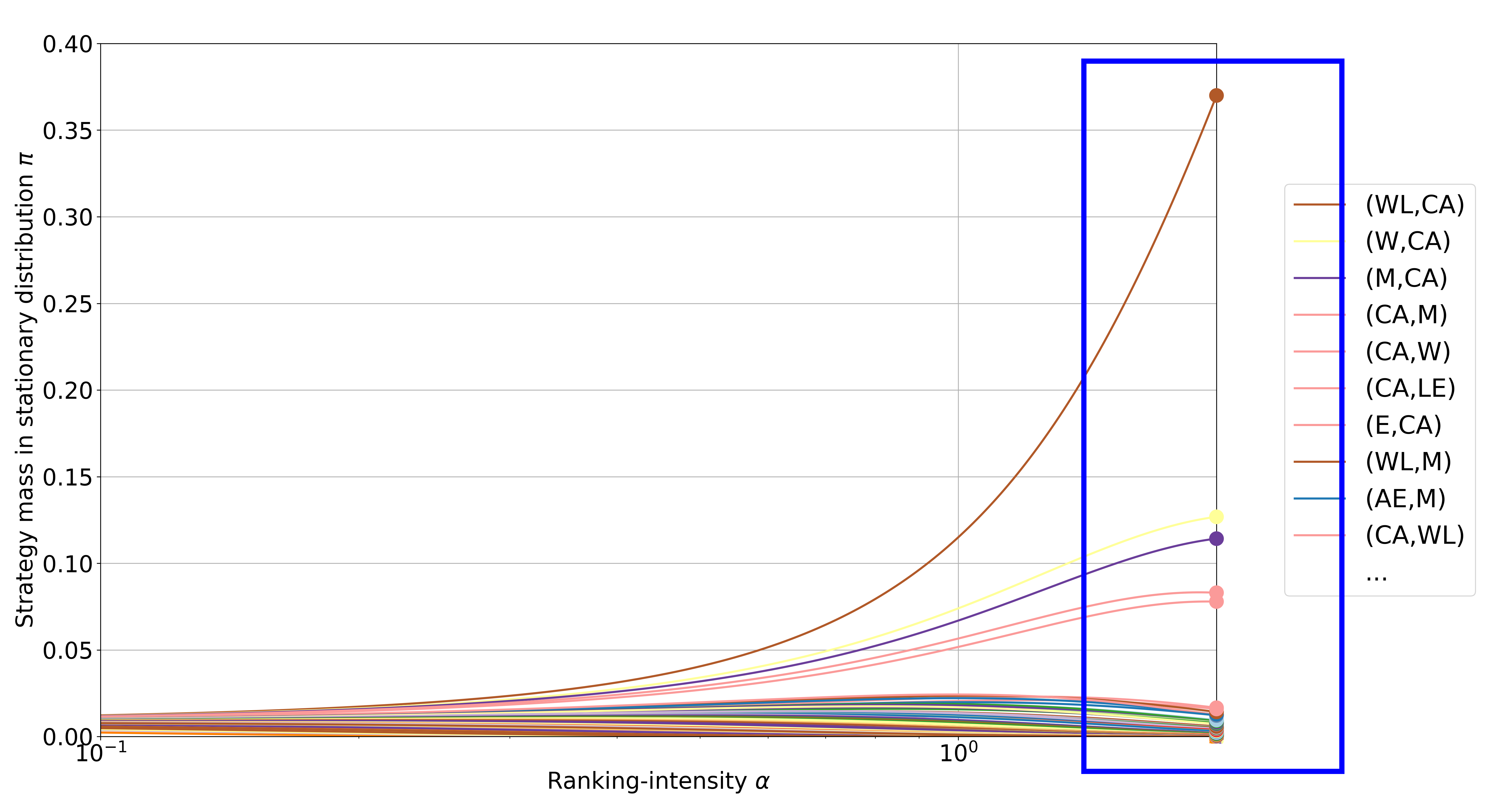}
    \caption{Mass across $\alpha \in [0.1, 3]$}
    \label{fig:alpha_x_pi_3}
  \end{subfigure}
  \hfill
  \begin{subfigure}[b]{0.49\linewidth}
    \includegraphics[width=\linewidth]{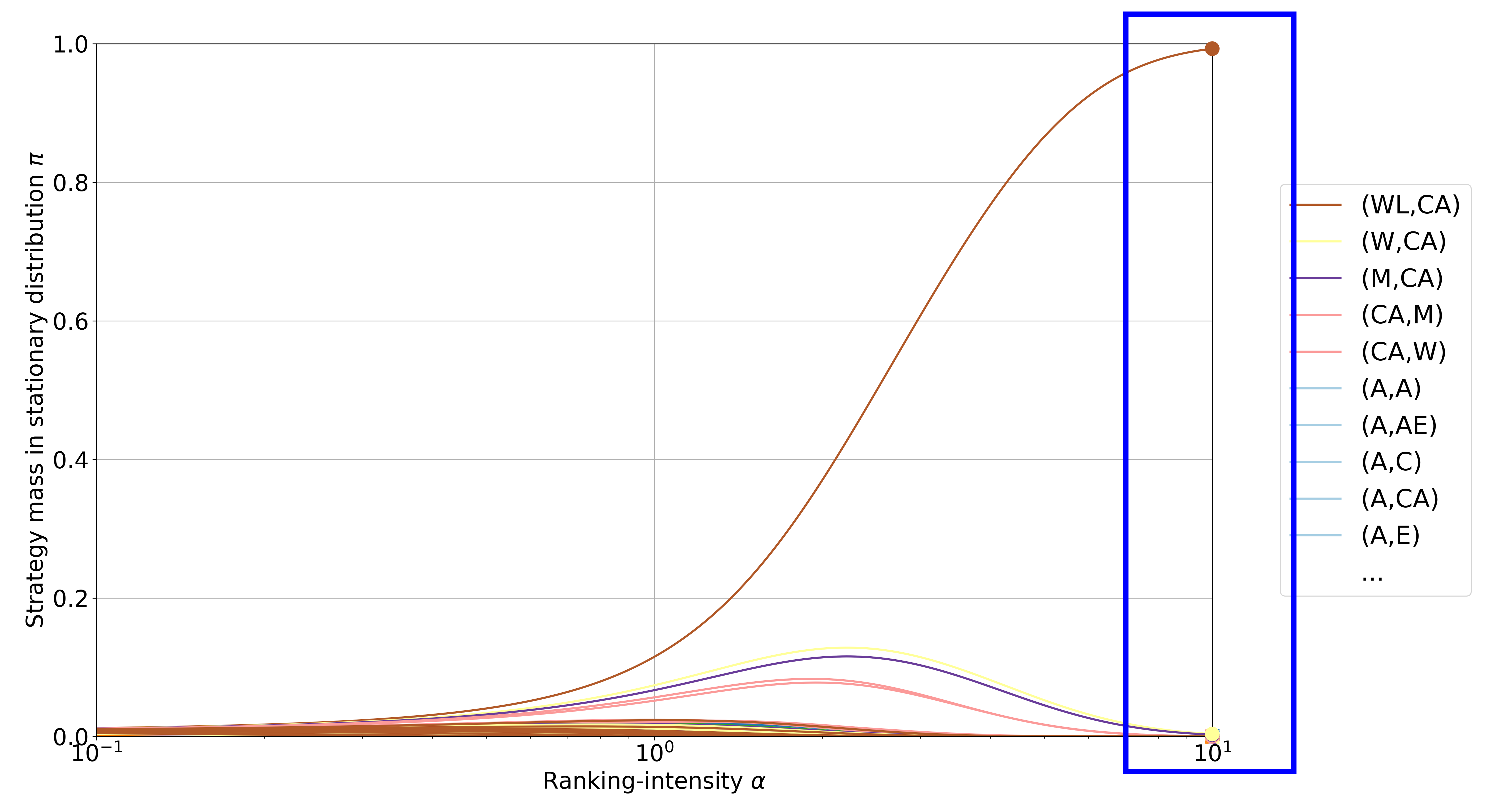}
    \caption{Mass across $\alpha \in [0.1, 10]$}
    \label{fig:alpha_x_pi_10}
  \end{subfigure}
  \caption{Effect of ranking intensity $\alpha$ on strategy profile mass in the stationary distribution $\pi$. The order of profiles in the legend matches the mass for maximal $\alpha$.}
  \label{fig:alpha_x_pi}
  \Description{Effect of ranking intensity $\alpha$ on strategy profile mass in the stationary distribution $\pi$. The order of profiles in the legend matches the mass for maximal $\alpha$.}
\end{figure*}

The response graph describes the dynamics of the strategy profiles in the empirical game. One prominent feature is the primary component of the MCC: the profile (WL, CA). This profile, indicated by a dark blue color, has multiple graph edges leading to it, while none from it, indicating that strategies in this profile are non-transient. This is further supported by the large fixation probabilities along the edges. A particularly prominent example is the cluster (CA, LE)-(CA, M)-(CA, W), which consists of three strongly connected profiles, indicating that once a player adopts one of these profiles, they will likely remain within their cluster. These components reflect stable regions in the game’s strategy dynamics, where transitions between profiles become locked into a cycle.

To further investigate the effect of $\alpha$ on profile dominance, we plotted the stationary distribution $\pi$ across all $\alpha$ values used in the experiments, for the top-performing strategy profiles (see Figure~\ref{fig:alpha_x_pi}). This visualization--also part of $\mathcal{D}$--helps us understand how the stationary distribution changes as the selection intensity increases. The x-axis represents the different $\alpha$ values, ranging from $0.1$ to $3$ in Figure~\ref{fig:alpha_x_pi_3}, and from $0.1$ to $10$ in Figure~\ref{fig:alpha_x_pi_10}, while the y-axis in both figures shows the mass of each strategy profile in the stationary distribution $\pi$. As $\alpha$ increases, the distribution converges, and the selection process stabilizes. The final mass distributions are highlighted in boxed regions. The legend at the right displays the top-performing joint strategies, with the stronger ones at the top.

We plot two such graphs to observe how the mass of strategy profiles is distributed in the MCCs across different $\alpha$ values. In the stationary distribution resulting from a bigger $\alpha$, the dominant strategy profile (WL, CA) in the MCC achieves a mass of 1, with all other profiles dropping to 0. This is clearly illustrated in the second plot (see Figure~\ref{fig:alpha_x_pi_10}). However, regarding the mass distribution for a smaller range of $\alpha$, depicted n the first plot, the game has not yet converged to the final MCC.


\section{Related work}
Evaluation and ranking of learned multi-agent strategies is mainly based on game theoretic concepts, while computational social choice has been proposed, as well \cite{lanctot2023evaluating}.
A predominant approach is the Elo rating system used to evaluate and rank agents that learn through reinforcement learning (\cite{elo1978rating}, \cite{silver2016mastering}, \cite{doi:10.1126/science.aar6404}, \cite{mnih2015human}). Elo estimates the probability an agent to win another agent. Although it was designed specifically to rank players in the two-player, symmetric constant-sum game, it has been widely applied to other domains. Recently, it has been used for the evaluation of large language models \cite{10.5555/3666122.3668142}. However, Elo cannot model intransitive relationships \cite{DBLP:conf/nips/BalduzziTPG18}, as those in the Rock-Paper-Scissors game, in real-world games \cite{NEURIPS2020_ca172e96}, and in our game coloring setting. In addition, incorrectness issues have been reported in transitive settings \cite{pmlr-v206-bertrand23a}. Nash-based evaluation methods, such as Nash Averaging, can be applied in two-player, zero-sum settings (\cite{DBLP:conf/nips/BalduzziTPG18} \cite{10.5555/3237383.3237402}), but these are not more generally applicable as the Nash equilibrium is intractable to compute and select \cite{10.1145/1132516.1132527}.

Here, our focus is on (many) agents' long-term interactions in general-sum dynamic settings: Agents need to rank their policy profiles to decide their non-transient individual policy profile, accounting for long-term interactions and payoffs. 

Empirical Game Theory Analysis (EGTA), deploying empirical or meta-games \cite{10.1007/11575726_8} \cite{tuyls2007evolutionary},  \cite{gerald2002analyzing}, \cite{wellman2006methods}, can be used to evaluate learning agents that interact in large-scale multiagent systems \cite{omidshafiei2019alpharank}, \cite{tuyls2018generalised} \cite{tuyls2018symmetric}. In our case, aiming at strategy profile rankings, we define the empirical game strategies based on agents' styles of play and train policies realizing these strategies. Then the empirical payoff matrix is estimated by means of game-playing simulations, and is used by the $\alpha$-Rank method to compute strategy profile rankings. $\alpha$-Rank applies to many-player, general-sum games \cite{omidshafiei2019alpharank}.



\section{Conclusions}

In this study, we developed a methodology for identifying strong joint-strategies in dynamic multi-agent games, accounting for stability and performance, using the \emph{$\alpha$-Rank} evolutionary algorithm. This methodology is applied to a stochastic version of the \emph{Graph Coloring Problem}, where players collaborate to color a graph while ensuring that neighboring vertices are assigned different colors. According to the methodology, we first transformed the game into its empirical form by defining styles of play. We then designed and trained Deep Q-Learning policy models that realize these styles of play in the underlying game, and run simulations to generate the empirical payoff matrix. Applying \emph{$\alpha$-Rank} to this matrix results in a unique stationary distribution over strategy profiles, which defines the empirical game's MCC. \emph{$\alpha$-Rank} not only helped us identify stable strategy profiles resistant to changes, but also provided a descriptive framework for understanding why certain profiles prevail in the long run, based on the underlying dynamics of the game. Through this approach, we successfully described a concise methodology for evaluating and ranking agents' joint policies, considering their long-term interactions in dynamic settings, while also explaining how strategy profiles are defined within the MCC.

Future work involves (a) applying the methodology in more complex and large-scale settings, accounting for strategy profiles of multiple stakeholders that may collaborate and/or compete, (b) using machine learning methods to identify different styles of play from demonstrations and specifying the empirical game, (c) exploring advanced models able to adapt their strategies based on observed behaviors based on the behavior of co-players, and (d) applying the methodology into real-world settings where agents need to align with human preferences in dynamic settings.


\newpage

\bibliographystyle{ACM-Reference-Format} 
\bibliography{refs}

\newpage

\appendix

\onecolumn

\section{Empirical Payoff Matrix}
\label{appendix:A}

This is the empirical payoff matrix derived from simulations of the \emph{Graph Coloring Game} using policies trained to adhere to specific styles of play. Each entry in the matrix represents the payoffs of strategies in the corresponding profile, with the first value indicating the payoff of the row strategy and the second value of the column player. The Nash equilibria are highlighted in bold, while nine of the top-ranked strategy profiles in the MCC are shaded in gray.
\begin{table*}[h!]
\centering
\resizebox{\textwidth}{!}{%
\begin{tabular}{lccccccccccc}
    \hline
        & A & AE & C & CA & E & I & L & LE & M & W & WL \\
    \hline
    A  & (3.12, 3.11) & (3.15, 3.16) & (3.17, 3.17) & (3.14, 3.17) & (3.16, 3.17) & (3.16, 3.15) & (3.22, 3.13) & (3.19, 3.16) & (3.15, 3.18) & (3.16, 3.17) & (3.21, 3.18) \\
    
    AE & (3.17, 3.17) & (3.11, 3.11) & (3.18, 3.17) & (3.15, 3.17) & (3.17, 3.16) & (3.19, 3.16) & (3.23, 3.12) & (3.19, 3.16) & (3.15, 3.18) & (3.17, 3.17) & (3.20, 3.16) \\
    
    C  & (3.17, 3.16) & (3.16, 3.17) & (3.10, 3.10) & (3.14, 3.17) & (3.15, 3.15) & (3.18, 3.15) & (3.22, 3.12) & (3.17, 3.14) & (3.14, 3.17) & (3.17, 3.16) & (3.20, 3.17) \\
    
    CA & (3.17, 3.15) & (3.17, 3.15) & (3.17, 3.14) & (3.11, 3.11) & (3.18, 3.15) & (3.18, 3.14) & (3.24, 3.13) & \cellcolor{gray!16}(3.21, 3.16) & \cellcolor{gray!16}(3.16, 3.16) & \cellcolor{gray!16}(3.19, 3.16) & (3.22, 3.15) \\
    
    E  & (3.15, 3.16) & (3.16, 3.16) & (3.15, 3.16) & (3.15, 3.17) & (3.10, 3.10) & (3.18, 3.16) & (3.22, 3.12) & (3.19, 3.14) & (3.15, 3.17) & (3.16, 3.17) & (3.19, 3.17) \\
    
    I  & (3.14, 3.16) & (3.16, 3.18) & (3.16, 3.18) & (3.15, 3.19) & (3.16, 3.17) & (3.12, 3.12) & (3.22, 3.14) & (3.18, 3.16) & (3.14, 3.19) & (3.16, 3.18) & (3.19, 3.18) \\

    L  & (3.14, 3.22) & (3.11, 3.22) & (3.12, 3.22) & (3.13, 3.23) & (3.12, 3.22) & (3.13, 3.22) & (3.12, 3.12) & (3.14, 3.20) & (3.11, 3.21) & (3.14, 3.23) & \cellcolor{red!16}\textbf{(3.15, 3.21)} \\
    
    LE & (3.15, 3.19) & (3.14, 3.18) & (3.14, 3.18) & (3.15, 3.21) & (3.15, 3.19) & (3.16, 3.17) & (3.20, 3.14) & (3.11, 3.11) & (3.14, 3.22) & (3.15, 3.18) & (3.18, 3.19) \\
    
    M  & (3.17, 3.14) & (3.17, 3.15) & (3.17, 3.15) & \cellcolor{gray!16}(3.16, 3.17) & (3.16, 3.14) & (3.18, 3.14) & (3.23, 3.11) & (3.20, 3.14) & (3.06, 3.08) & (3.18, 3.15) & (3.20, 3.16) \\
    
    W  & (3.17, 3.17) & (3.17, 3.18) & (3.16, 3.18) & \cellcolor{gray!16}(3.16, 3.20) & (3.17, 3.17) & (3.18, 3.16) & (3.21, 3.13) & (3.18, 3.15) & (3.15, 3.18) & (3.08, 3.09) & (3.19, 3.15) \\
 
    WL & (3.17, 3.20) & (3.17, 3.19) & (3.17, 3.19) & \cellcolor{gray!16}(3.17, 3.22) & (3.17, 3.19) & (3.18, 3.19) & \cellcolor{red!16}\textbf{(3.21, 3.15)} & (3.19, 3.17) & (3.16, 3.20) & (3.16, 3.19) & (3.13, 3.13) \\
    \hline
\end{tabular}
}
\vspace{0.5em}
\caption{Empirical Payoff Matrix for the Graph Coloring Game}
\label{tab:gcg_payoff_matrix}
\end{table*}
%

\clearpage
\section{Response Graph}
\label{appendix:B}

These are four response graphs that illustrate the dynamics of strategy profiles in the empirical \emph{Graph Coloring Game} for different $\alpha$ values. Each node in the graph represents a unique strategy profile in the MCC, while the edges indicate transitions between them. The values on the edges show the fixation probabilities normalized by the neutral fixation probability, denoted as $\rho_m$. The nodes and edges are color-coded. Darker blue nodes represent more strong joint profiles, while lighter blue nodes represent transient ones. Similarly, bold arrows suggest a strong advantage in shifting between the nodes, whereas faint ones suggest less of an advantage. 
\begin{figure*}[h]
  \centering
  \begin{subfigure}[b]{0.48\linewidth}
    \includegraphics[width=\linewidth]{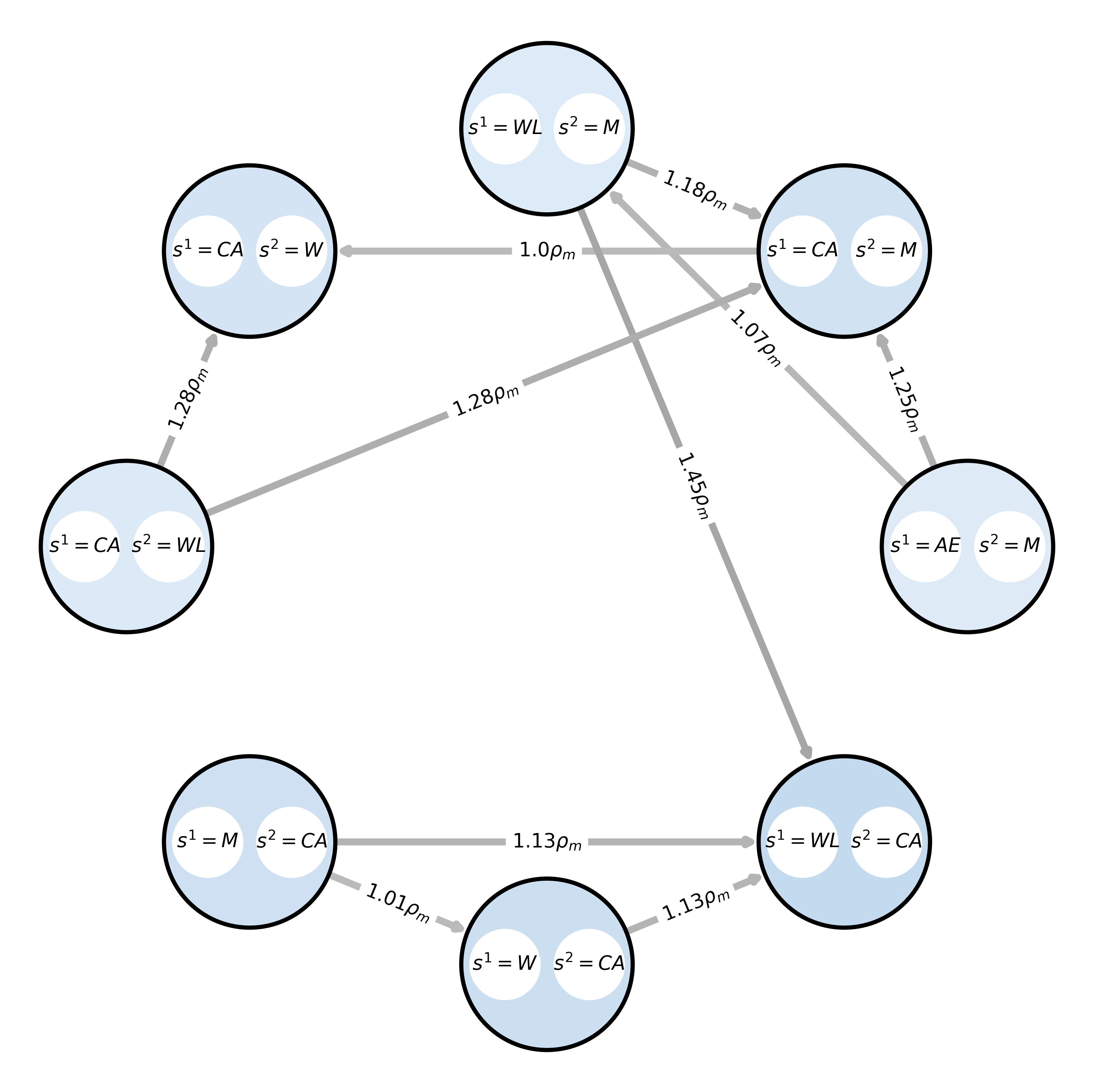}
    \caption{$alpha=0.4$}
    \label{fig:response_graph_0.4}
  \end{subfigure}
  \hfill
  \begin{subfigure}[b]{0.48\linewidth}
    \includegraphics[width=\linewidth]{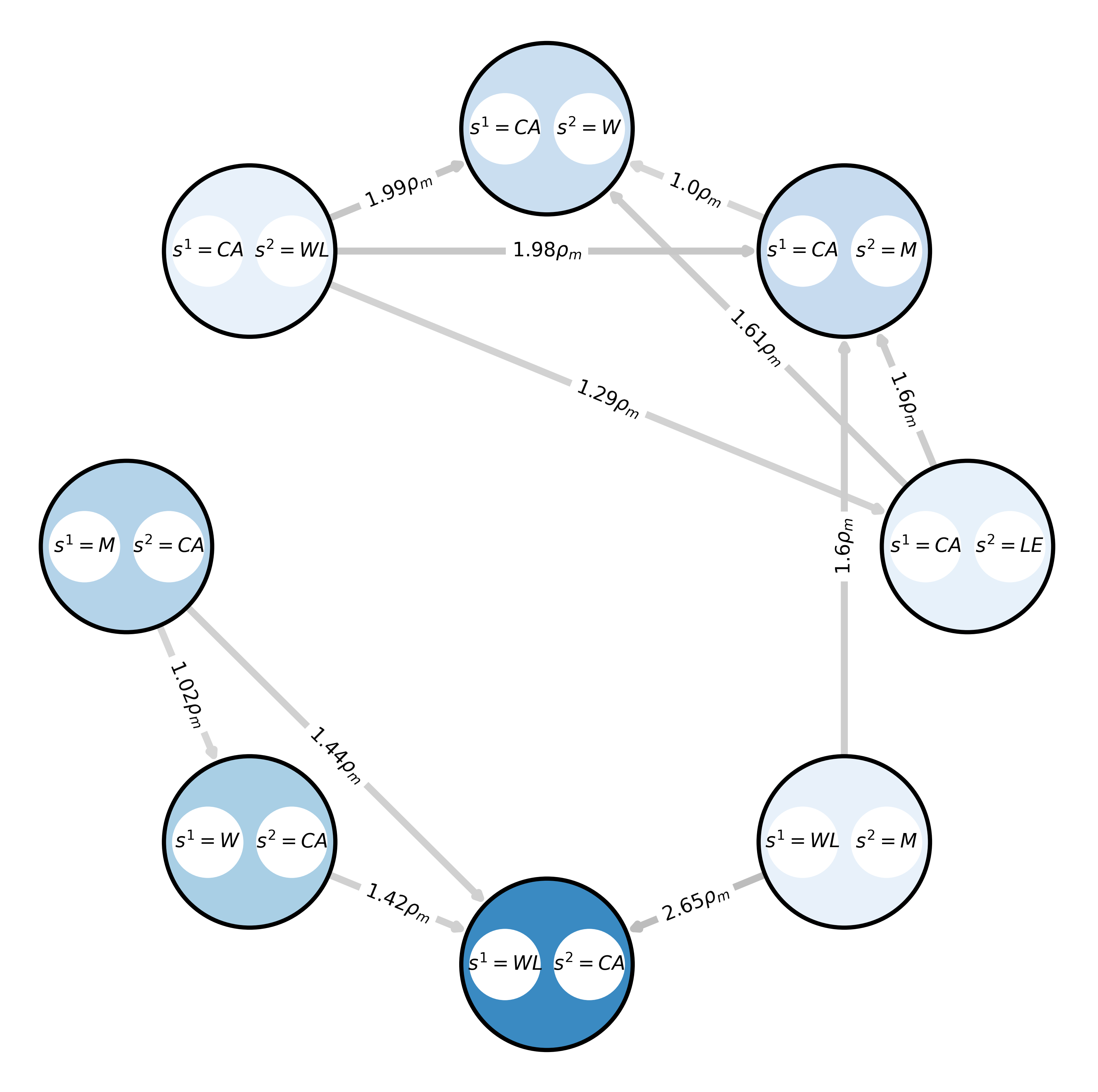}
    \caption{$alpha=1.3$}
  \end{subfigure}

  \begin{subfigure}[b]{0.48\linewidth}
    \includegraphics[width=\linewidth]{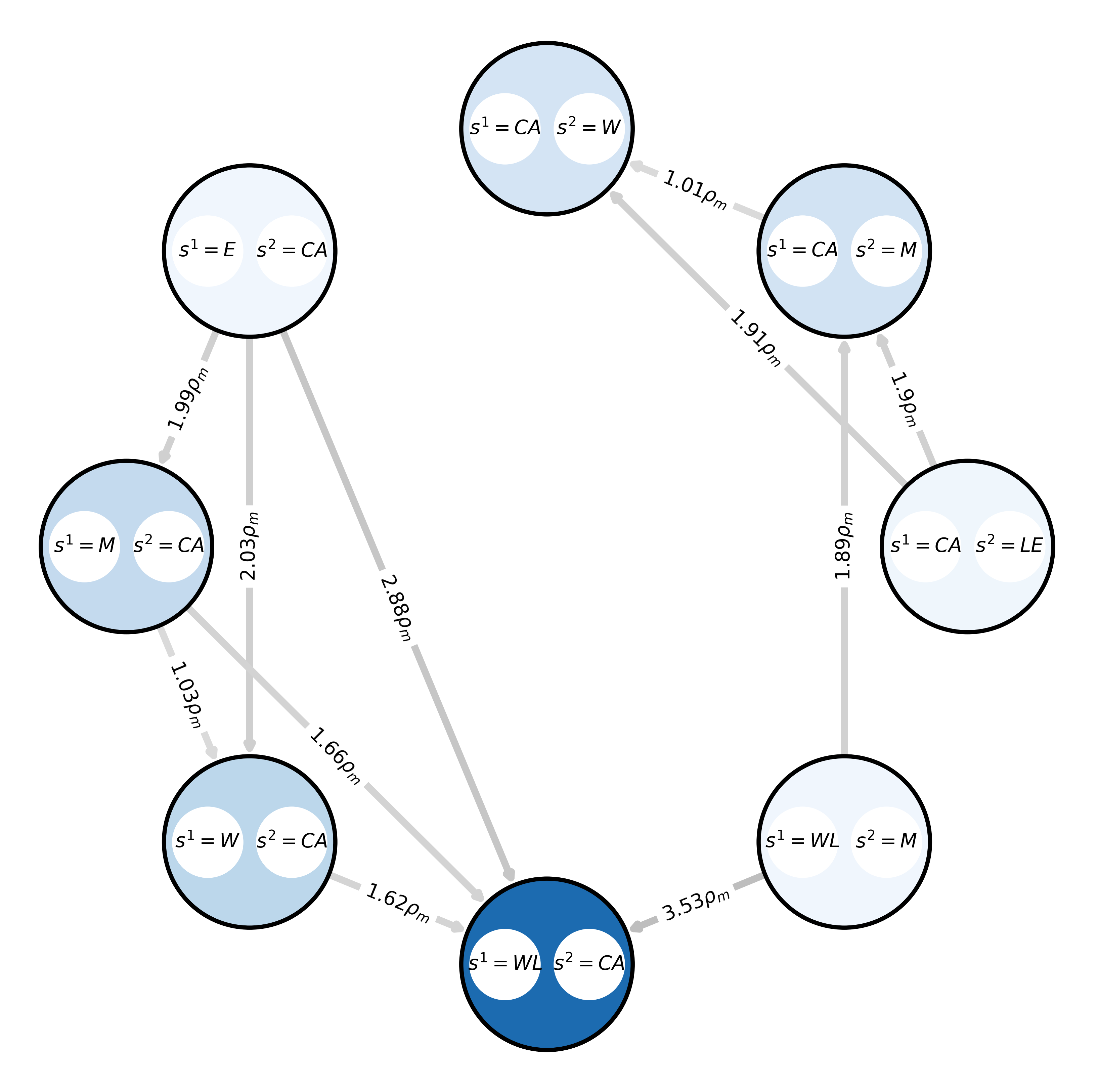}
   \caption{$alpha=1.9$}
  \end{subfigure}
  \hfill
  \begin{subfigure}[b]{0.48\linewidth}
    \includegraphics[width=\linewidth]{rg_6.4.png}
    \caption{$alpha=6.4$}
  \end{subfigure}

  \caption{Response graphs of strategy profiles' dynamics.}
  \label{fig:response_graphs}
  \Description{Response graphs of strategy profiles' dynamics.}
\end{figure*}
%

\clearpage
\section{Experimenting and aggregating rankings using different configurations}
\subsection{Experiments in additional game configurations}
\label{appendix:C1}

\begin{figure}[h]
  \centering
   \begin{subfigure}{0.40\textwidth}
      \centering
      \includegraphics[width=\textwidth]{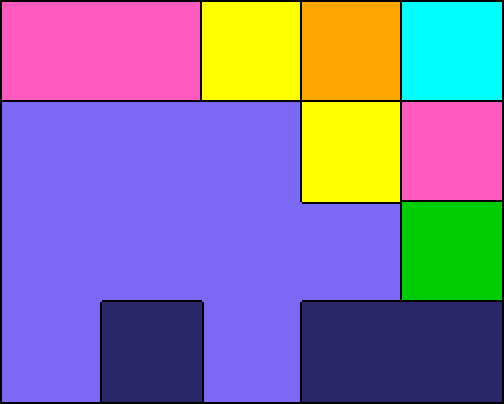}
      \caption{}
      \label{fig:grid_1}
   \end{subfigure}
   \hspace{0.05\textwidth}
   \begin{subfigure}{0.40\textwidth}
      \centering
      \includegraphics[width=\textwidth]{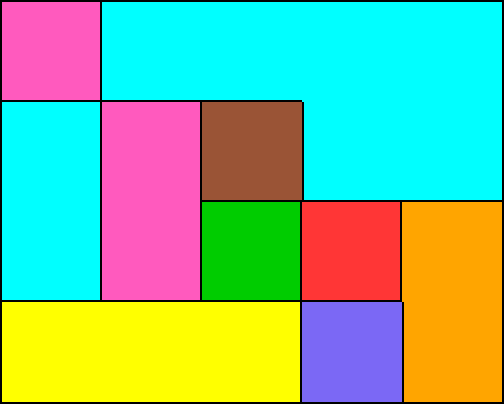}
      \caption{}
      \label{fig:grid_2}
   \end{subfigure}
   \caption{The $4 \times 5$ grid environments used for the two additional experiments on the graph coloring game.}
   \label{fig:grids}
   \Description{The $4 \times 5$ grid environments used for the two additional experiments on the graph coloring game.}
 \end{figure}

Here we show experiments and results with two additional $4 \times 5$ with 10 blocks game configurations, shown in Figure~\ref{fig:grids}. The strategies, the underlying policy model architecture and the setup for training the policy models realizing these strategies are as described in the article. According to the proposed method, for each configuration, we generate the empirical payoff matrix by simulating each strategy profile across multiple games. Finally, we apply the $\alpha$-Rank method to evaluate the performance of strategy profiles.

 As observed from the rankings in Tables~\ref{tab:ranking_table_1} and \ref{tab:ranking_table_2}, the strategy profiles that prevail vary between the two configurations. For the first configuration, the top-ranked profile is (C, W) with a score of 0.18 (Table~\ref{tab:ranking_table_1}), while for the second configuration the best-performing profile is (W, CA) with a score of 0.34 (Table~\ref{tab:ranking_table_2}). First, while the top-ranked profiles differ between configurations, both include strategies with cool (C) and warm (W) preferences. This is reasonable, as these preferences inherently guarantee minimal overlap in color selection, reducing conflicts and improving overall performance. The difference in highly ranked profiles can be explained by the blocks with the largest number of neighbors in the different game configurations. Having many blocks with many neighbors, the configuration results to a dense graph: a strategy preferring a large block coloring difficulty (i.e., ``A") is rewarded proportionally to the number of chosen block neighbors, being indifferent in colors chosen. Combined with ``C" (i.e., ``CA") it prefers to color the chosen blocks with cool colors. Therefore, in dense graph configurations, ``CA" will chose blocks that constrain the choice of colors for many other blocks, and will color them with cool colors: This allows ``W" strategy to be more rewarding when coloring the blocks not chosen by ``CA".
 

 Another pair of strategies that appears among the top 6 best-performing profiles in both configurations are the lazy (L) and ambitious (A) preferences. These strategies complement each other naturally, as a lazy player tends to choose blocks with fewer neighbors, while an ambitious player does not prefer the coloring of such blocks. This combination minimizes the likelihood of selecting the same block, thus preventing penalties due to overlapping choices.

 \begin{table}[h]
     \centering
     \begin{subtable}{0.45\textwidth}
         \centering
         \begin{tabular}{lcc}
             \hline
             \textbf{Profile} & \textbf{Rank} & \textbf{Score} \\
             \hline
             (C, W) & 1 & 0.18 \\
             (M, CA) & 2 & 0.11 \\
             (M, AE) & 3 & 0.10 \\
             (M, C) & 4 & 0.06 \\
             (WL, AE) & 5 & 0.05 \\
             (CA, M) & 6 & 0.05 \\
             \hline
         \end{tabular}
         \caption{}
         \label{tab:ranking_table_1}
     \end{subtable}
     \hspace*{0pt}
     \begin{subtable}{0.45\textwidth}
         \centering
         \begin{tabular}{lcc}
             \hline
             \textbf{Profile} & \textbf{Rank} & \textbf{Score} \\
             \hline
             (W, CA) & 1 & 0.34 \\
             (W, E) & 2 & 0.07 \\
             (M, W) & 3 & 0.05 \\
             (WL, CA) & 4 & 0.04 \\
             (E, W) & 5 & 0.04 \\
             (W, M) & 6 & 0.03 \\
             \hline
         \end{tabular}
         \caption{}
         \label{tab:ranking_table_2}
     \end{subtable}
     \caption{Strategy profiles' rankings for $\alpha=2$}
     \label{tab:ranking_tables}
 \end{table}

\begin{figure}[h]
  \centering
   \begin{subfigure}{0.47\textwidth}
      \centering
       \includegraphics[width=1\linewidth]{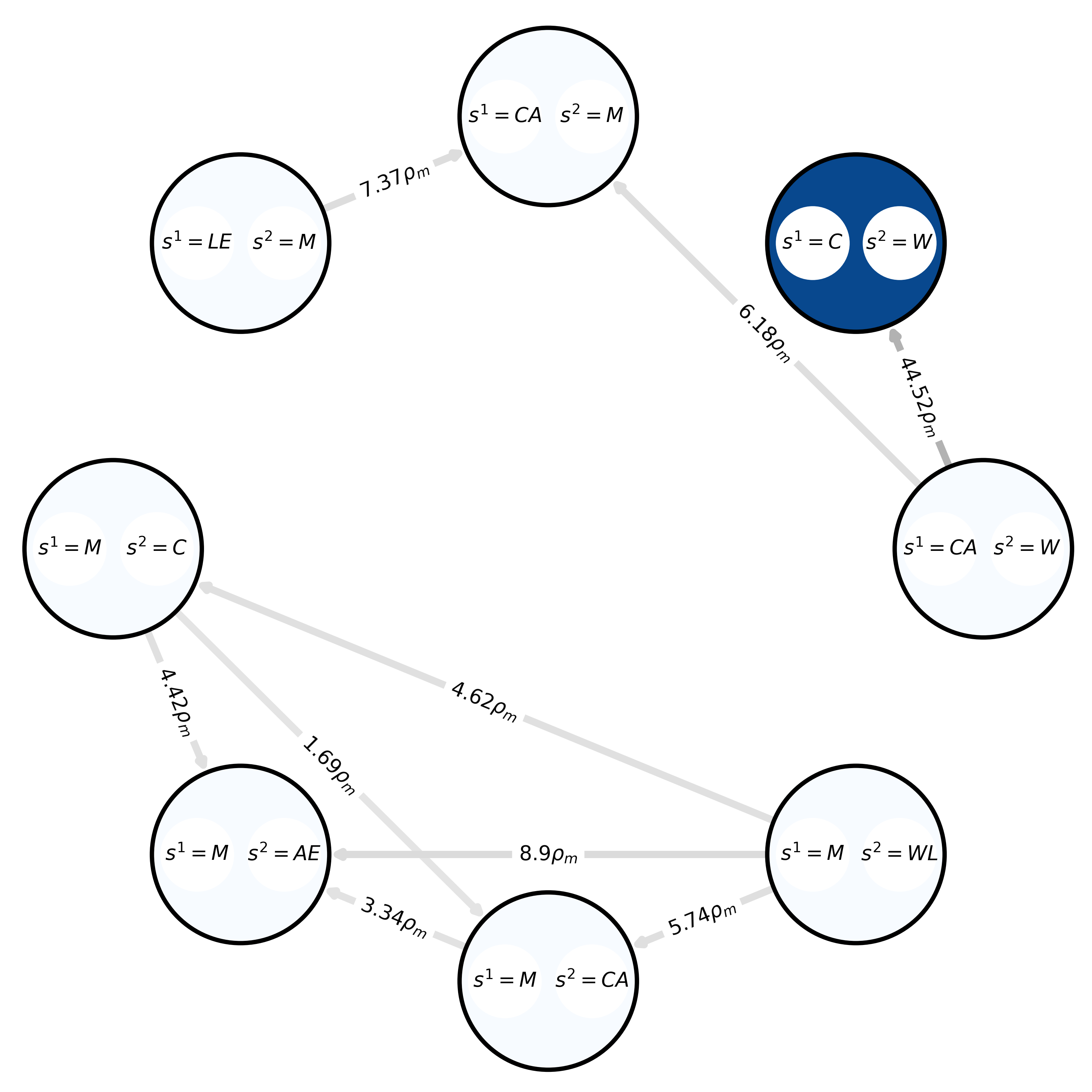}
       \caption{}
       \label{fig:response_graph_1}
   \end{subfigure}
   \hspace{0.05\textwidth}
   \begin{subfigure}{0.47\textwidth}
      \centering
       \includegraphics[width=1\linewidth]{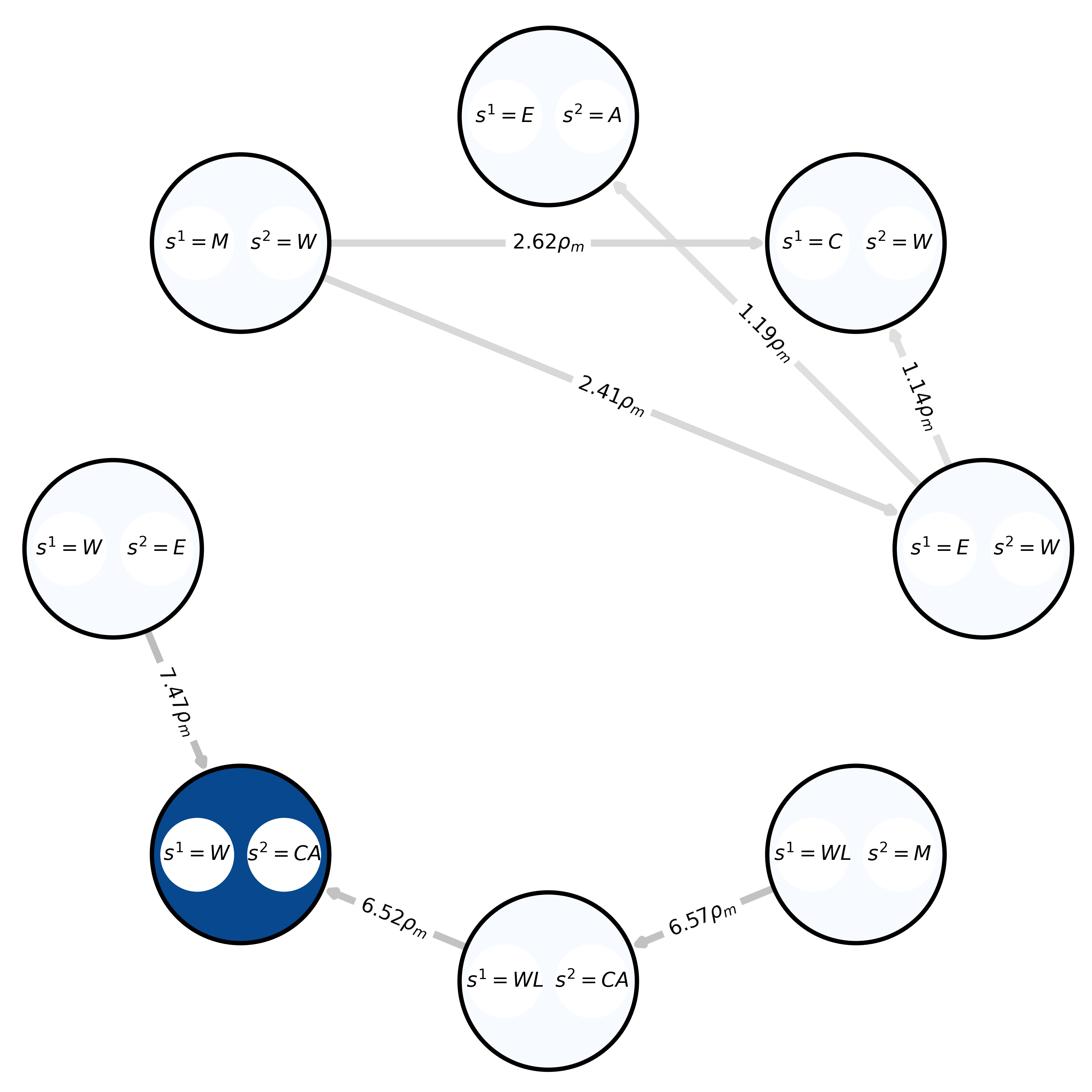}
       \caption{}
       \label{fig:response_graph_2}
   \end{subfigure}
   \caption{Response graphs for $\alpha=6.4$.}
   \label{fig:response_graphs_appendix}
   \Description{Response graphs for $\alpha=6.4$.}
 \end{figure}

 The response graphs for the two configurations, as shown in Figure~\ref{fig:response_graphs_appendix}, illustrate the dynamics of strategy profiles in the empirical game. In both cases, the response graphs reveal two strongly connected components, and the top-ranked profiles form the primary components of the MCC: (C, W) in Figure~\ref{fig:response_graph_1} and (W, CA) in Figure~\ref{fig:response_graph_2}. 

\begin{figure*}[h]
   \centering
   \begin{subfigure}[b]{0.49\linewidth}
     \includegraphics[width=\linewidth]{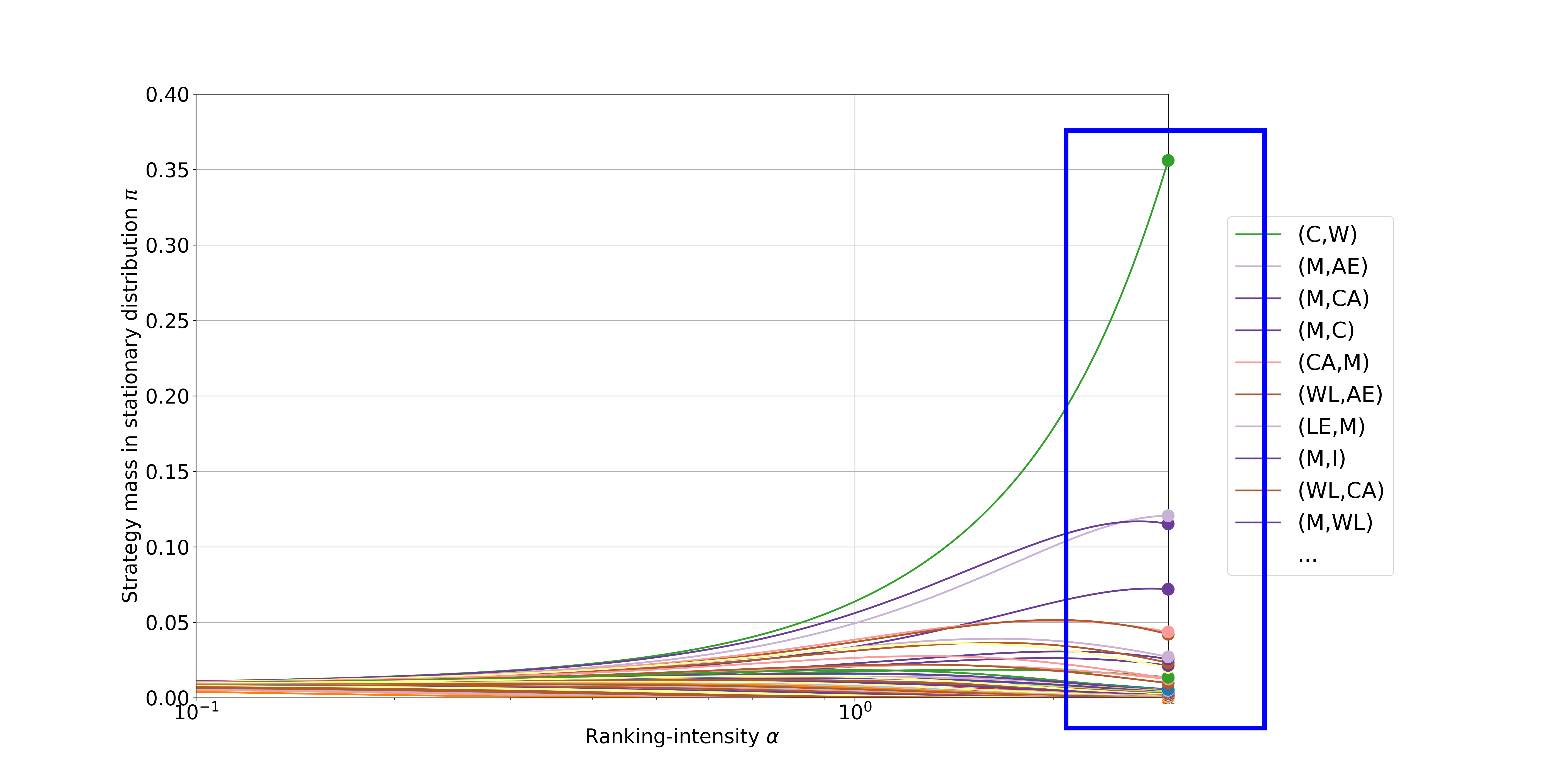}
     \caption{}
     \label{fig:alpha_x_pi_3_1}
   \end{subfigure}
   \hfill
   \begin{subfigure}[b]{0.49\linewidth}
     \includegraphics[width=\linewidth]{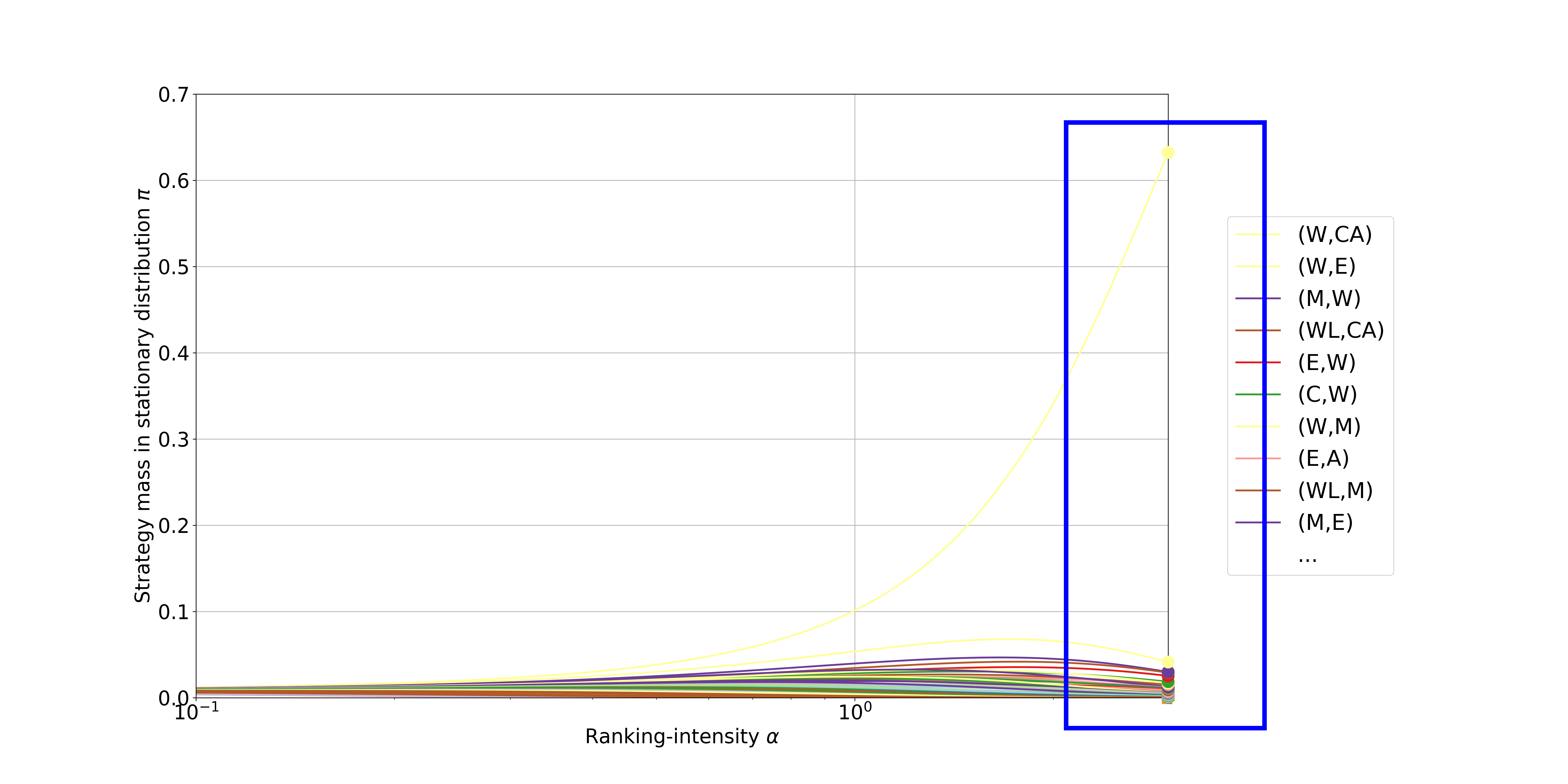}
     \caption{}
     \label{fig:alpha_x_pi_3_2}
   \end{subfigure}
   \caption{Effect of ranking intensity $\alpha \in [0.1, 3]$ on strategy profile mass in the stationary distribution $\pi$.}
   \label{fig:alpha_x_pis}
   \Description{Effect of ranking intensity $\alpha \in [0.1, 3]$ on strategy profile mass in the stationary distribution $\pi$.}
 \end{figure*}

 To further investigate the effect of $\alpha$ on the ranking of profiles, we plot the stationary distribution $\pi$ across all $\alpha$ values, ranging from 0.1 to 3, per experiment (Figure~\ref{fig:alpha_x_pis}).

\newpage
\subsection{Cross-configurations policy rankings}
\label{appendix:C2}
To determine which strategy profiles perform best across different graph configurations, we aggregate data from all three experiments (the graph configuration shown in Figure \ref{fig:grid_and_graph} and the two configurations in Figure \ref{fig:grids}) and applied the $\alpha$-Rank method. 
Data aggregation here involves taking the empirical payoff matrices produced in the different configurations and computing cell-wise average values. This results into a new ``average" empirical payoff matrix. 
Ranking policy profiles in this way reduces the risk of overfitting to any specific configuration and provides more generalized and reliable rankings. Having done so, the rankings for the top 6 strategy profiles are presented in Table~\ref{tab:ranking_table_3}. 
However, aggregating empirical payoff matrices for different configurations may not be of value in cases these configurations differ largely. This is an aspect that needs more study.

 \begin{table}[h]
   \centering
   \begin{tabular}{lcc}
       \hline
       \textbf{Profile} & \textbf{Rank} & \textbf{Score} \\
       \hline
       (M, CA) & 1 & 0.31 \\
       (W, CA) & 2 & 0.09 \\
       (CA, M) & 3 & 0.09 \\
       (C, W) & 4 & 0.05 \\
       (CA, W) & 5 & 0.04 \\
       (WL, CA) & 6 & 0.03 \\
       \hline
   \end{tabular}
   \vspace{0.5em}
   \caption{Averaged strategy profiles' rankings for $\alpha=2$}
   \label{tab:ranking_table_3}
 \end{table}

 The aggregated results highlight that (M, CA) ranks first with a significant score of 0.31. This agrees with the arguments for the rankings in the different configurations discussed in \hyperref[appendix:C1]{Appendix C1}, given also the fact that a minimalistic strategy is rewarded by choosing colors already in use, proportionally to their frequency of use.  We conjecture that this ranking shows that the minimalistic strategy learns to choose blocks that are not adjacent to those chosen by the CA strategy, using colors similar to those used by the CA strategy.
 It is interesting to note that profiles (W, CA), in Table~\ref{tab:ranking_table_2}, (C, W) in Table~\ref{tab:ranking_table_1}, and (WL, CA) in Table~\ref{tab:ranking_table}, are ranked second, fourth, and sixth, respectively, in the aggregated data. This suggests that their performance remains strong across all three configurations, unlike many other strategy profiles. 

 \begin{figure*}[h]
   \centering
   \includegraphics[width=0.85\linewidth]{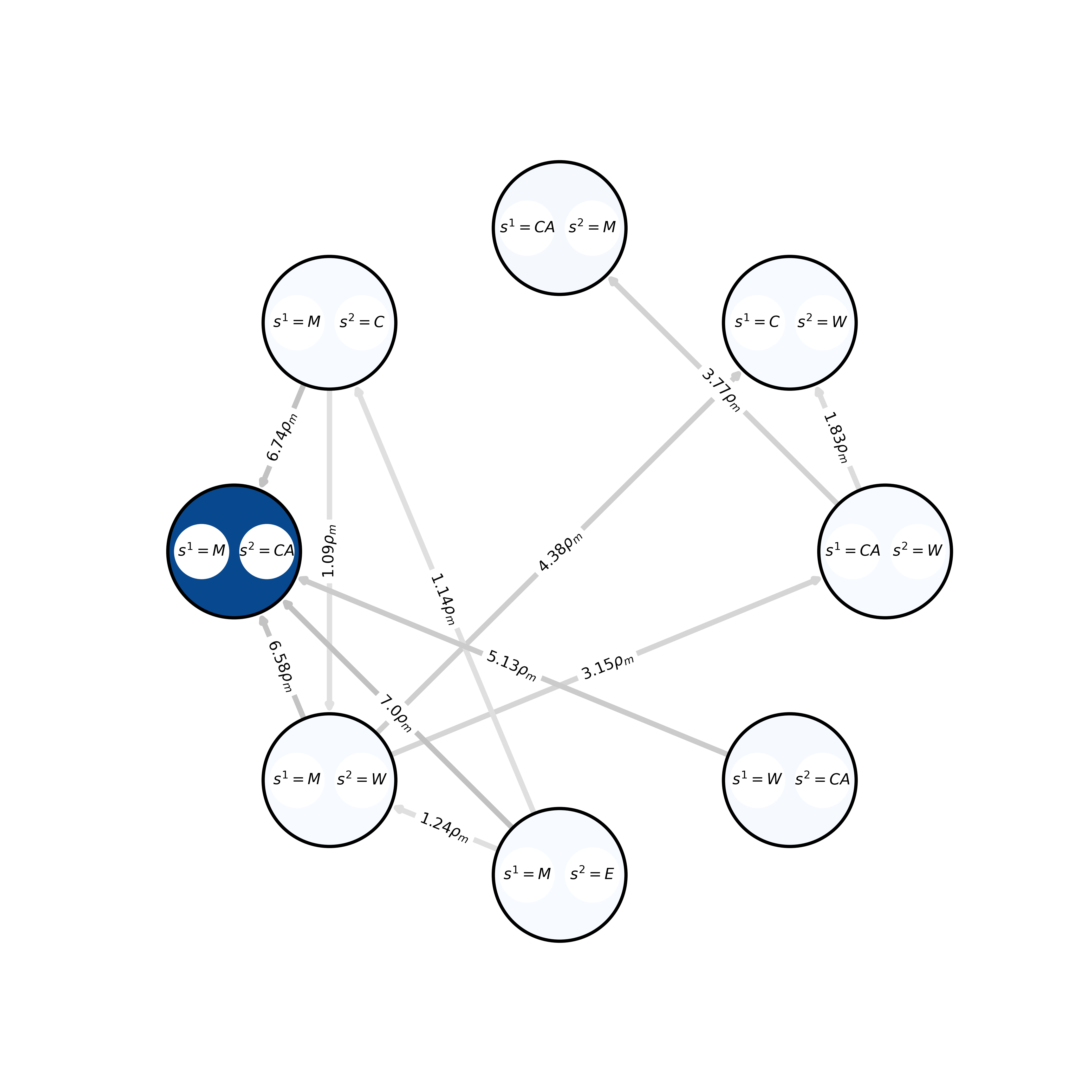}
   \caption{Response graph for $\alpha=6.4$}
   \label{fig:response_graph_6.4_3}
   \Description{Response graph for $\alpha=6.4$}
 \end{figure*}

\begin{figure*}[h]
   \centering
   \includegraphics[width=0.85\linewidth]{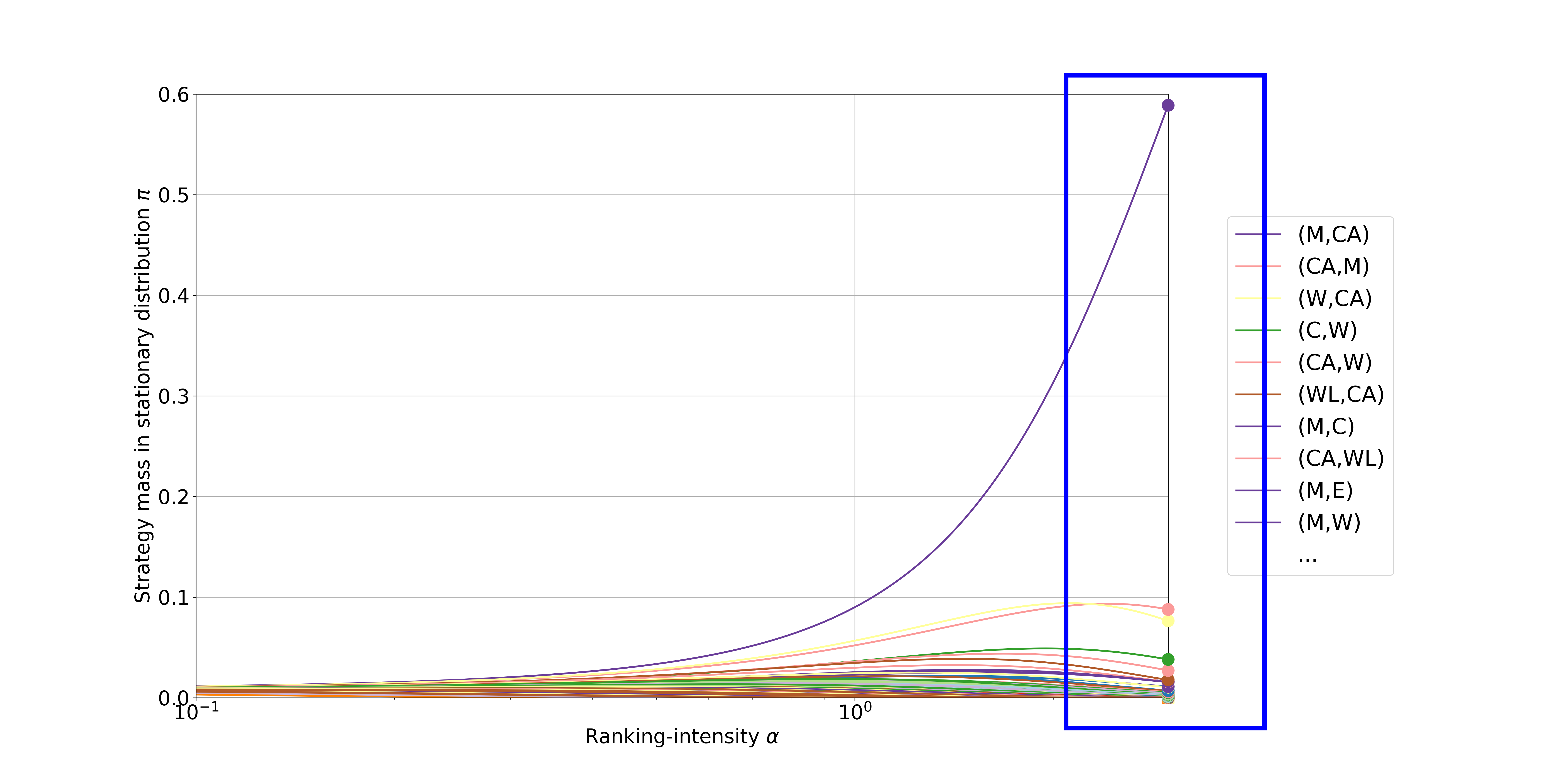}
   \caption{Effect of ranking intensity $\alpha \in [0.1, 3]$ on strategy profile mass in the stationary distribution $\pi$.}
   \label{fig:alpha_x_pi_3_3}
   \Description{Effect of ranking intensity $\alpha \in [0.1, 3]$ on strategy profile mass in the stationary distribution $\pi$.}
 \end{figure*}

 The response graph for the aggregated results (Figure~\ref{fig:response_graph_6.4_3}) illustrates the overall dynamics of the strategy profiles. The top-ranked profile, (M, CA), is highlighted in dark blue, signifying that once the game reaches this MCC, it is unlikely to transition to any other profile. The graph also shows that all 8 profiles form a large connected cluster, with varying degrees of connection between them. 
To further investigate the effect of $\alpha$ on profile dominance, we plotted the stationary distribution $\pi$ across all $\alpha$ values, ranging from 0.1 to 3 (Figure~\ref{fig:alpha_x_pi_3_3}).

 %

 \clearpage
 \section{The Role of Convolutions in Policy Training}
 \label{appendix:D}

\begin{figure}[h]
   \centering
   \includegraphics[width=0.6\linewidth]{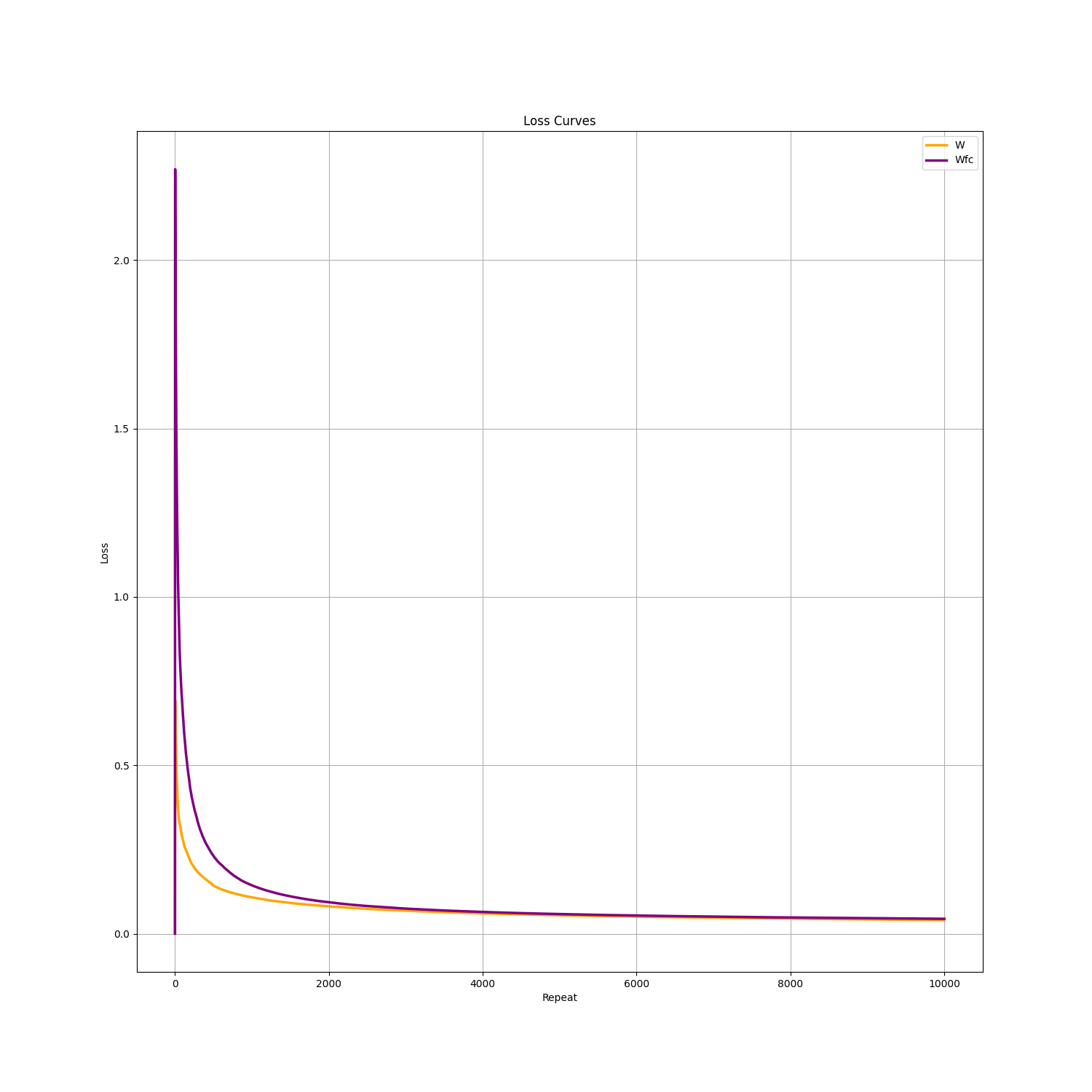}
   \caption{Training efficiency comparison: W uses convolutional layers (orange) and Wfc fully connected layers (purple).}
   \label{fig:conv_vs_noconv}
   \Description{Training efficiency comparison: W uses convolutional layers (orange) and Wfc fully connected layers (purple).}
 \end{figure}

We conjecture that the grid-like structure of the environment makes convolutions a natural choice for the policy model architecture: Since the grid includes spatial relationships between blocks, local dependencies, and configurations with multiple blocks, convolutions enable the policy model to efficiently capture these spatial patterns. Thus, incorporating convolutional layers, policy models are enabled to converge more effectively. To validate this claim, we performed an experiment that compares the training of policies with and without convolutional layers. The results are presented in Figure~\ref{fig:conv_vs_noconv}.

Specifically, we trained a policy for the ``warm" strategy using the model architecture with convolutional layers (W), and compared its performance to a ``warm" policy trained with a model of fully connected layers (Wfc). Both policies are trained on the same graph configuration, and their performance is evaluated based on the loss curves. The results show that the policy with convolutional layers converges more quickly, suggesting that convolutions enable the model to learn the spatial relationships between blocks more efficiently. It must be noted that the Wfc curve picks at 2.27, while the W at 0.89. While both policies eventually reach the same level of performance (with the loss after 10,000 epochs converging to the same low point), the model with the convolutional layers achieves this more effectively.

\end{document}